\documentclass[apj]{emulateapj}
\usepackage{graphics}
\usepackage{color}
\usepackage{rotating}

\shorttitle{Small-scale coronal heating events with Hi-C}
\shortauthors{R\'egnier, Alexander, Walsh et al.}

\begin{document}

\title{Sparkling EUV bright dots observed with Hi-C}
\author{S. R\'egnier\altaffilmark{1}}
\author{C. E. Alexander\altaffilmark{1}}
\author{R. W. Walsh\altaffilmark{1}}
\author{A.~R. Winebarger\altaffilmark{2}} 
\author{J. Cirtain\altaffilmark{2}}
\author{L. Golub\altaffilmark{3}}
\author{K.~E. Korreck\altaffilmark{3}}
\author{N. Mitchell\altaffilmark{4}}
\author{S. Platt\altaffilmark{4}}
\author{M. Weber\altaffilmark{3}}
\author{B. De Pontieu\altaffilmark{5}}
\author{A. Title\altaffilmark{5}}
\author{K. Kobayashi\altaffilmark{6}}
\author{S. Kuzin\altaffilmark{7}}
\author{C.~E. DeForest\altaffilmark{8}}

\altaffiltext{1}{Jeremiah Horrocks Institute, University of Central Lancashire,
Preston, Lancashire, PR1 2HE, UK}
\altaffiltext{2}{NASA Marshall Space Flight Center, VP 62, Huntsville, AL 35812, USA}
\altaffiltext{3}{Harvard-Smithsonian Center for Astrophysics, 60 Garden Street,
Cambridge, MA 02138, USA}
\altaffiltext{4}{School of Computing, Engineering and Physical Sciences,
University of Central Lancashire, Preston, Lancashire, PR1 2HE, UK}
\altaffiltext{5}{Lockheed Martin Solar and Astrophysics Laboratory, 3251 Hanover
Street, Palo Alto, CA 94304, USA}
\altaffiltext{6}{Center for Space Plasma and Aeronomic Research, 320 Sparkman
Dr, Huntsville, AL 35805, USA}
\altaffiltext{7}{Lebedev Physical Institute, Russian Academy of Sciences,
Leninskii pr. 53, Moscow, 119991 Russia}
\altaffiltext{8}{Southwest Research Institute, 1050 Walnut Street Suite 300,
Boulder, CO 80302, USA}

\begin{abstract}
Observing the Sun at high time and spatial scales is a step towards
understanding the finest and fundamental scales of heating events in the solar
corona. The Hi-C instrument has provided the highest spatial and temporal
resolution images of the solar corona in the EUV wavelength range to date. Hi-C
observed an active region on 11 July 2012, which exhibits several interesting
features in the EUV line at 193\AA: one of them is the existence of short, small
brightenings ``sparkling" at the edge of the active region; we call these EUV
Bright Dots (EBDs). Individual EBDs have a characteristic duration of 25s with a
characteristic length of 680 km. These brightenings are not fully resolved by
the SDO/AIA instrument at the same wavelength, however, they can be identified
with respect to the Hi-C location of the EBDs. In addition, EBDs are seen in
other chromospheric/coronal channels of SDO/AIA suggesting a temperature between
0.5 and 1.5 MK. Based on their frequency in the Hi-C time series, we define
four different categories of EBDs: single peak, double peak, long duration, and
bursty EBDs. Based on a potential field extrapolation from an SDO/HMI
magnetogram, the EBDs appear at the footpoints of large-scale trans-equatorial
coronal loops. The Hi-C observations provide the first evidence of small-scale
EUV heating events at the base of these coronal loops, which have a free magnetic
energy of the order of 10$^{26}$ erg.
\end{abstract}


\section{Introduction}

A long standing problem in solar physics is how the solar corona is heated up to
a temperature of few million degrees while the photosphere is at about 6000 K.
Several models have been developed either to heat the coronal plasma locally or
to transport heat from the lower layers of the solar atmosphere into the corona
(see for instance review by \citeauthor{klim06} \citeyear{klim06}), however,
these models often lack an explanation for an average temperature of 1MK for
the entire corona and its sustainability during a solar cycle. This strongly
suggests that several mechanisms may be at play, each of them providing a
substantial amount of heat to the corona and, which can vary during the activity
cycle of the Sun. Those models also invoke smaller spatial and time scales than
the ones provided by the current observations. 

So, the critical issue is really to answer the questions: where is the source of
heat located? and where and how does the heating take place? Along a given
magnetic structure (often a coronal loop), three scenarios are often considered:
the release of energy is located (i) at the footpoints of the structure in the
photospheric layer which is the most favorable location owed to the importance
of the photospheric motions and the source of material and energy transported
from the convection zone, (ii) at the apex of the structure, and (iii) uniformly
along the structure. Those three scenarios imply that the heat is transported
along the magnetic structures, and that we actually observe loop-like structures
in the corona. In terms of observations, it has been difficult to disentangle
the actual location of the heat source. As an example, several authors using a
dataset obtained in soft X-rays by {\em Yohkoh}/SXT have reached an opposite
conclusion depending on how they dealt with the data reduction
\citep{mac00,pri00,reale02}. To determine the possible heating scenario, the
authors looked at the profile of temperature along the coronal loop and compared
with a theoretical model. These indirect methods were needed because of the
coarse spatial and time resolutions of {\em Yohkoh}/SXT (compared to what can be
achieved nowadays). In this paper, with the development of new high spatial and
time resolution instrumentation, we are looking for a direct evidence of the
source of heat deposition, observationally charaterised by a small, short
increase in intensity.   

Theoretically, the different heating locations (apex, uniform or footpoint)
arise from the idea that magnetic energy stored in the chromosphere/corona
produces small, short bursts converting the magnetic energy into thermal and/or
kinetic energy (rapidly converted into heat). This is the basic principle of
\citeauthor{par88}'s theory (\citeyear{par88}) for heating the solar corona by
nanoflares. A large number of those small events is needed to sustain and
uniformly distribute the heat in the entire solar corona. The mechanism
responsible for the release of magnetic energy could be, for instance, magnetic
reconnection, wave mode coupling or turbulence. As a large amount of the free
energy contained in the magnetic field structure of active regions is located at
the base of the corona or in the chromosphere \citep{reg07}, the most favorable
heating location is at the footpoints of coronal loops. For the quiet Sun, the
footpoints of the coronal strucutres are the most effected by the photospheric
motions such as granular motions which can easily produce
tangled/braided/twisted magnetic flux bundles. In this paper, we combine EUV
observations and a magnetic field model to determine if Parker's model is a
viable scenario, even if we cannot identify the mechanism(s) responsible for the
energy release.  

High-resolution ground-based instruments in the visible-wavelength domain 
allowed to observe new structures and phenomena in the lower layers of the solar
atmosphere, especially small-scale brightenings or events such as spicules or
Ellerman bombs \cite[see e.g.,][]{sch03,ber04,rut04,rou05,bec07}. In
conjunction, the investigation of the coronal plasma at EUV wavelengths is in
constant improvement in terms of spatial and temporal resolution. For instance,
the Atmospheric Imager Assembly (AIA; \citeauthor{lemen12} \citeyear{lemen12})
on board the Solar Dynamics Observatory (SDO; \citeauthor{pes12}
\citeyear{pes12}) is largely contributing to a new era of fast (12s) small-scale
(pixel size of 0.6\arcsec) observations of the solar corona always discovering
new phenomena. As a new step forward, the Hi-C instrument has provided
high-temporal (5.5s) and high-spatial (pixel size of 0.1\arcsec) resolution
observations of the EUV corona. Using this dataset, \citet{cirt13} have recently
shown the existence of small-scale braided structures within the active region,
and evidenced the magnetic reconnection process occurring within the braided
structures.   In addition, two recent articles relying on Hi-C data have
provided new insights in the existence of small-scale heating events.
\cite{tes13} have found that the active region moss is highly dynamic with a
variability on timescales of 15s, and the moss brightenings lead to an energy
release of 10$^{23}$ erg (order of magnitude consistent with the nanoflare
model). Similarly, \cite{win13} have investigated transient brightenings along
small-scale transition-region loops. It was found that the loops have a diameter
ranging from 0.9\arcsec\ to 1.1\arcsec\ with an average duration of 64s.
\cite{win13} have estimated that these brightenings  radiate an energy of about
10$^{24}$--10$^{25}$ erg again an energy estimate consistent with the nanoflare
model.

In this paper, we report on the existence of an additional heat source for the
coronal plasma: EUV, short-lived (25 s), small-scale ($<$1\arcsec) brightenings taking
place in the low layers of the solar atmosphere (log T = 5.5--6.5). We call
these brightenings EUV bright dots (EBDs). This new discovery is made possible
by the high spatial resolution and fast time cadence of the Hi-C data.

In Section~\ref{sec:obs}, after introducing the Hi-C instrument, we
characterize  the location, size, and thermal properties of EBDs as derived from
the Hi-C channel and six SDO/AIA channels. We thus study their time evolution in
the observed region (Section~\ref{sec:dyn}), as well as their magnetic
properties (Section~\ref{sec:mag}) relying on a potential field extrapolation of
one SDO/HMI magnetogram. In Section~\ref{sec:em_loci}, the possible temperature
of EBDs is investigated using the EM loci method. The conclusions are drawn in
Section~\ref{sec:concl}.  

\begin{figure}
\begin{center}
\includegraphics[width=1.\linewidth]{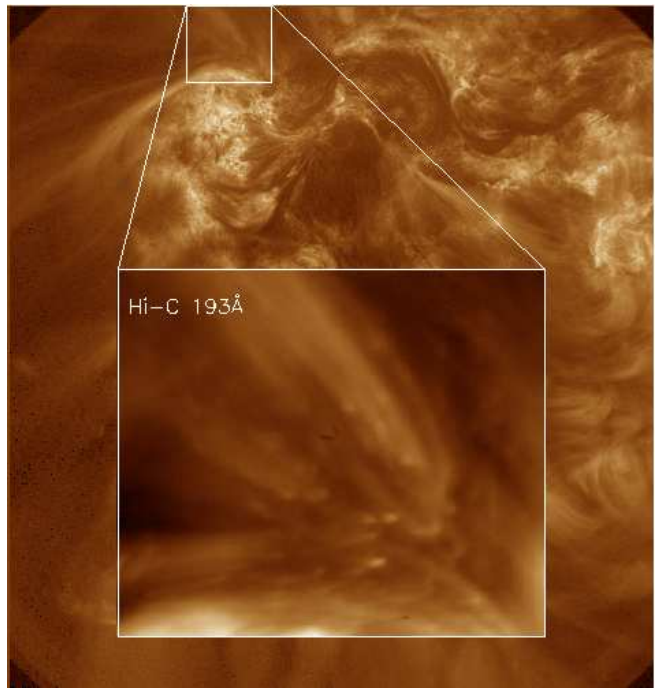}
\caption{Location of the region of interest containing the EBDs in the Hi-C FOV
recorded at 18:53:32 UT on 11 July 2012.}
\label{fig:hic_context}
\end{center}
\end{figure}
    
\section{Structure of EUV Bright dots}
\label{sec:obs}

	\subsection{The Hi-C Instrument}

The High-resolution Coronal (Hi-C) imager was launched on a sounding rocket on
July 11, 2012 \citep{kob13}. The Hi-C passband isolates a very narrow window in
EUV centered at 193 \AA, taking images with a pixel size of 0.1\arcsec. The
intensity line profile is dominated by an emission line of Fe {\sc xii} with a
peak temperature at 1.5 MK (log T = 6.17), and blended by some weak transition
region lines, as well as including 10-15 MK emission during large flares. The
main target of the Hi-C instrument was the complex system of active regions
located around the central meridian in the Southern hemisphere (active regions
NOAA 11519-21), and which indeed mostly contains the active region 11520
(AR11520). In Fig.~\ref{fig:hic_context}, we plot the full Hi-C field-of-view
(FOV) centered approximately at [-150, -281] arcseconds from the Sun center. We
extract the region of interest containing the EBDs discussed in this paper. The
size of the EBDs FOV is of approximately 35\arcsec$\times$30\arcsec.
Magnetically, AR11520 is a collection of 7-8 sunspots as can be seen in
Fig.~\ref{fig:bz} left. The Hi-C time series lasts during
200 s from 18:52:09 UT to 18:55:29 UT with a time cadence of 5.5 s (36 images).
	
	\subsection{Comparing Hi-C and SDO/AIA Observations}\label{sec:comp}

\begin{figure}[!h]
\begin{center}
\includegraphics[width=0.49\linewidth]{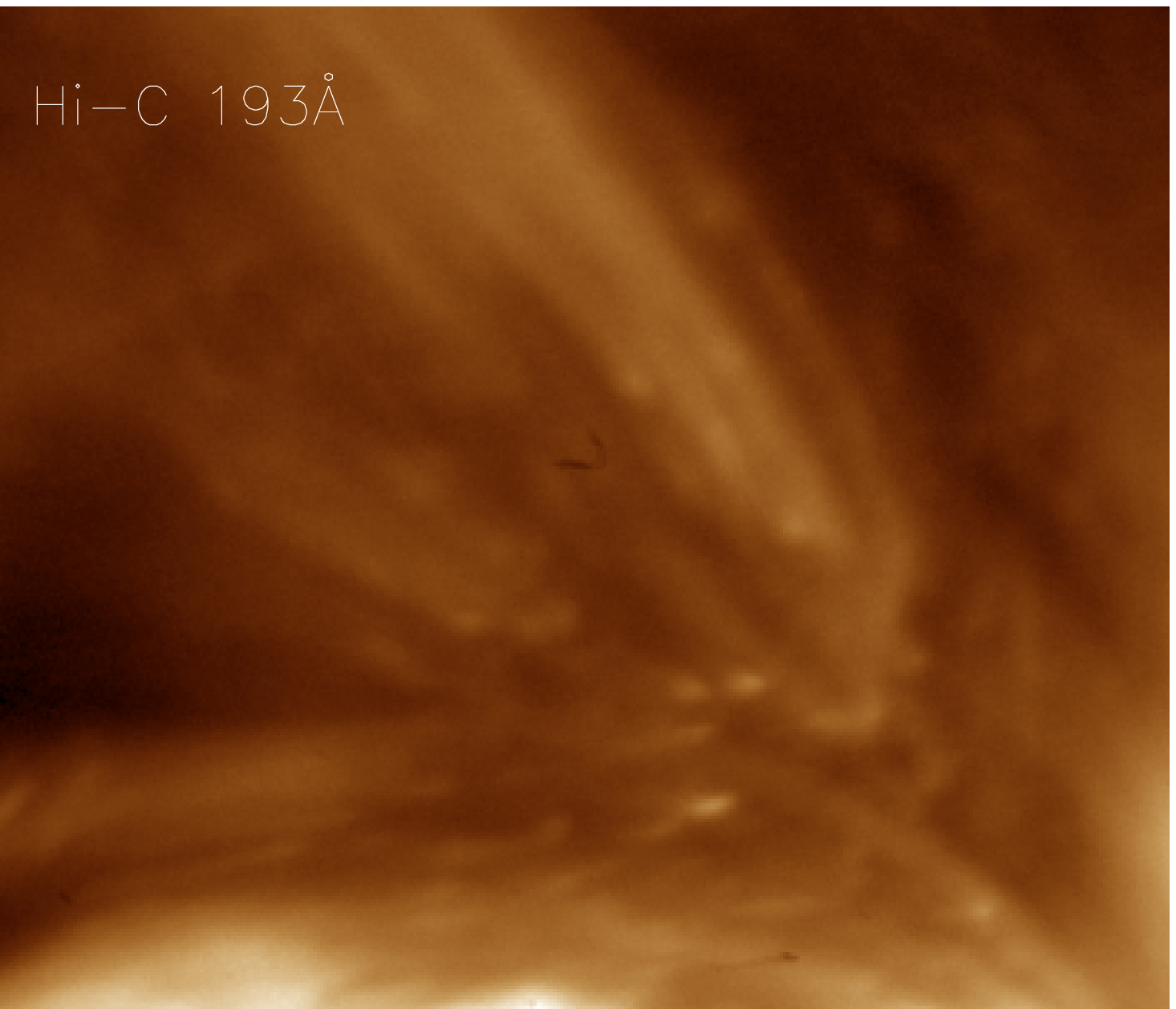}
\includegraphics[width=0.49\linewidth]{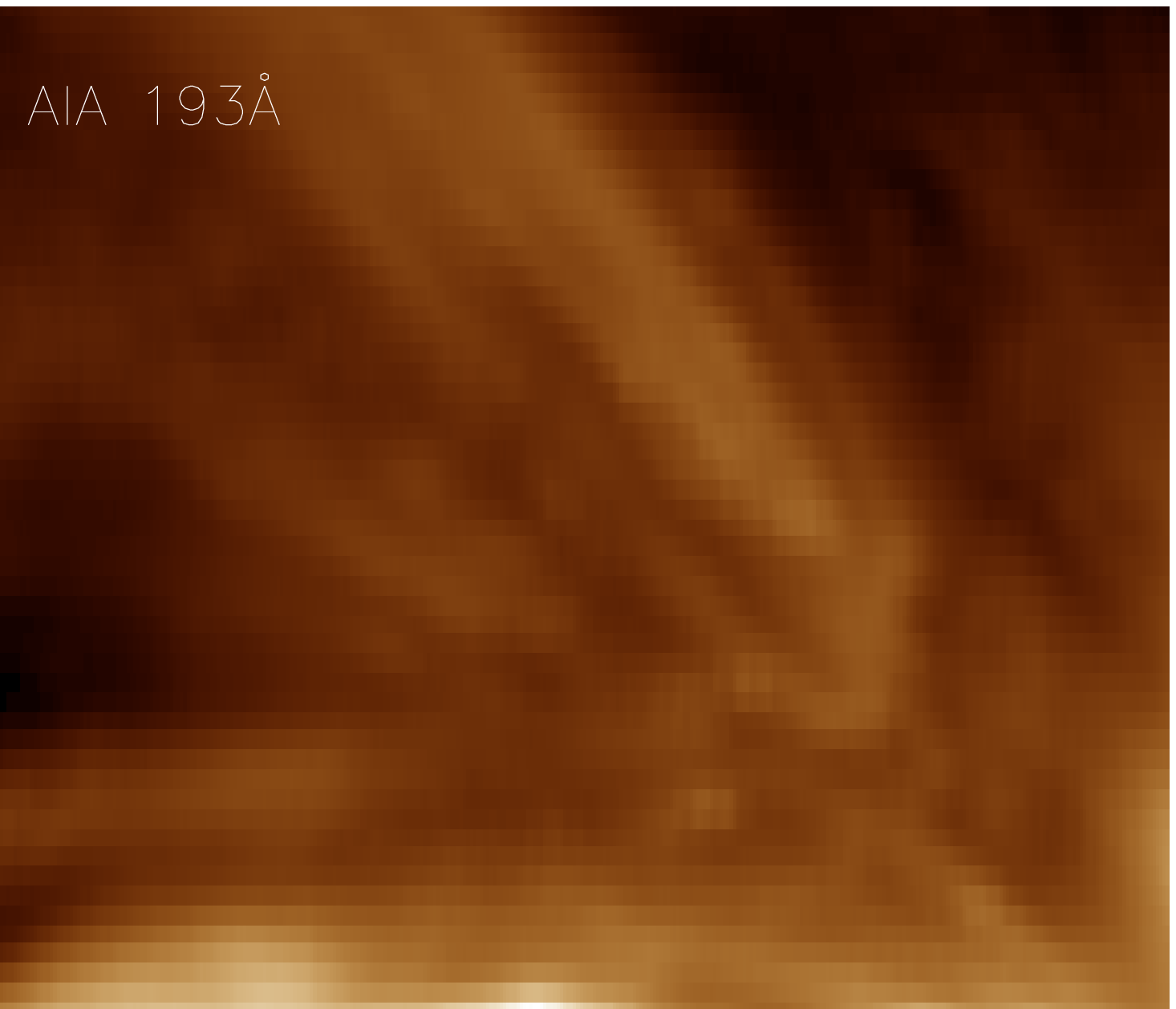}
\includegraphics[width=0.49\linewidth, bb= 0 0 420 360]{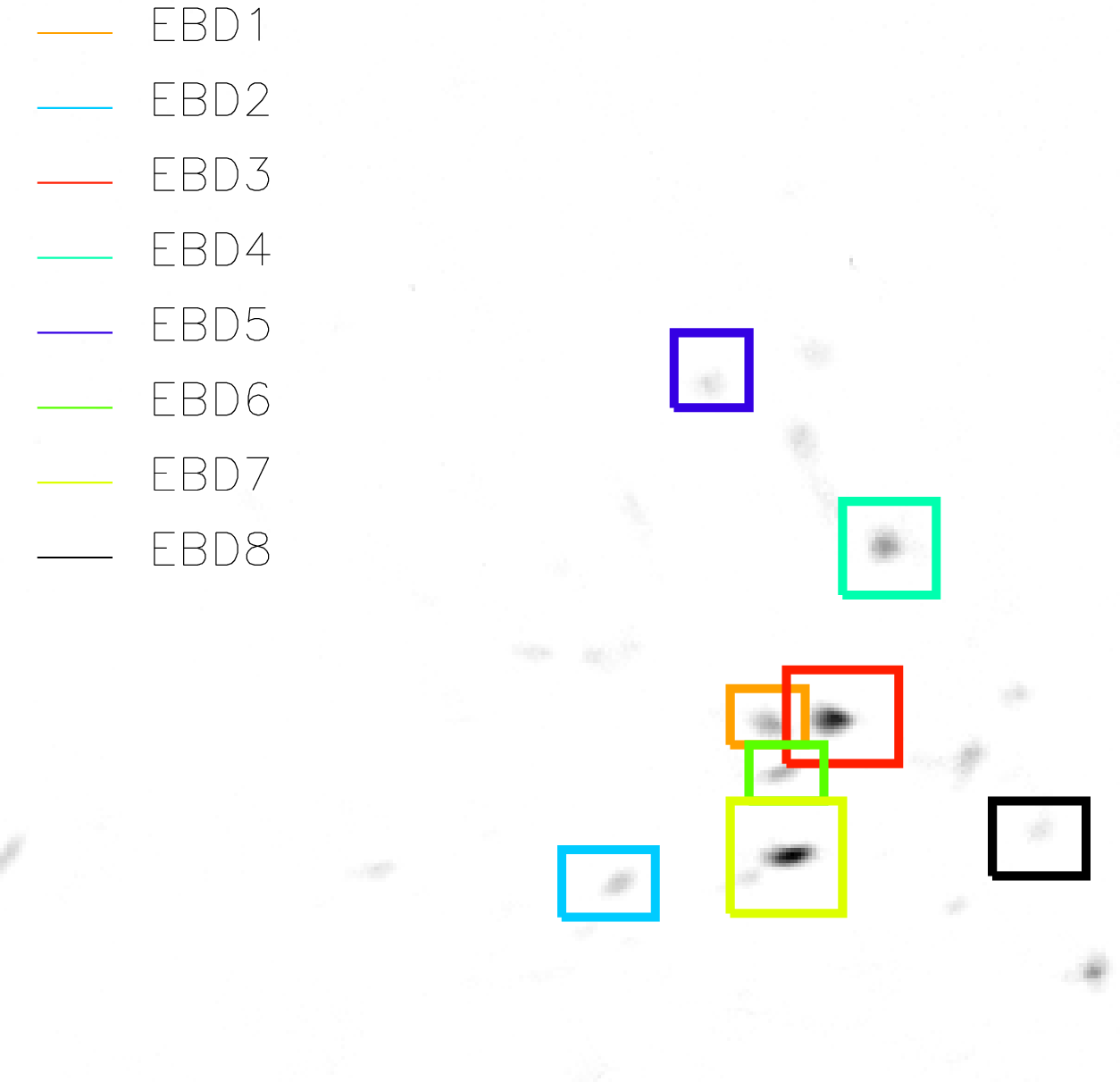}
\includegraphics[width=0.49\linewidth, bb= 0 0 420 360]{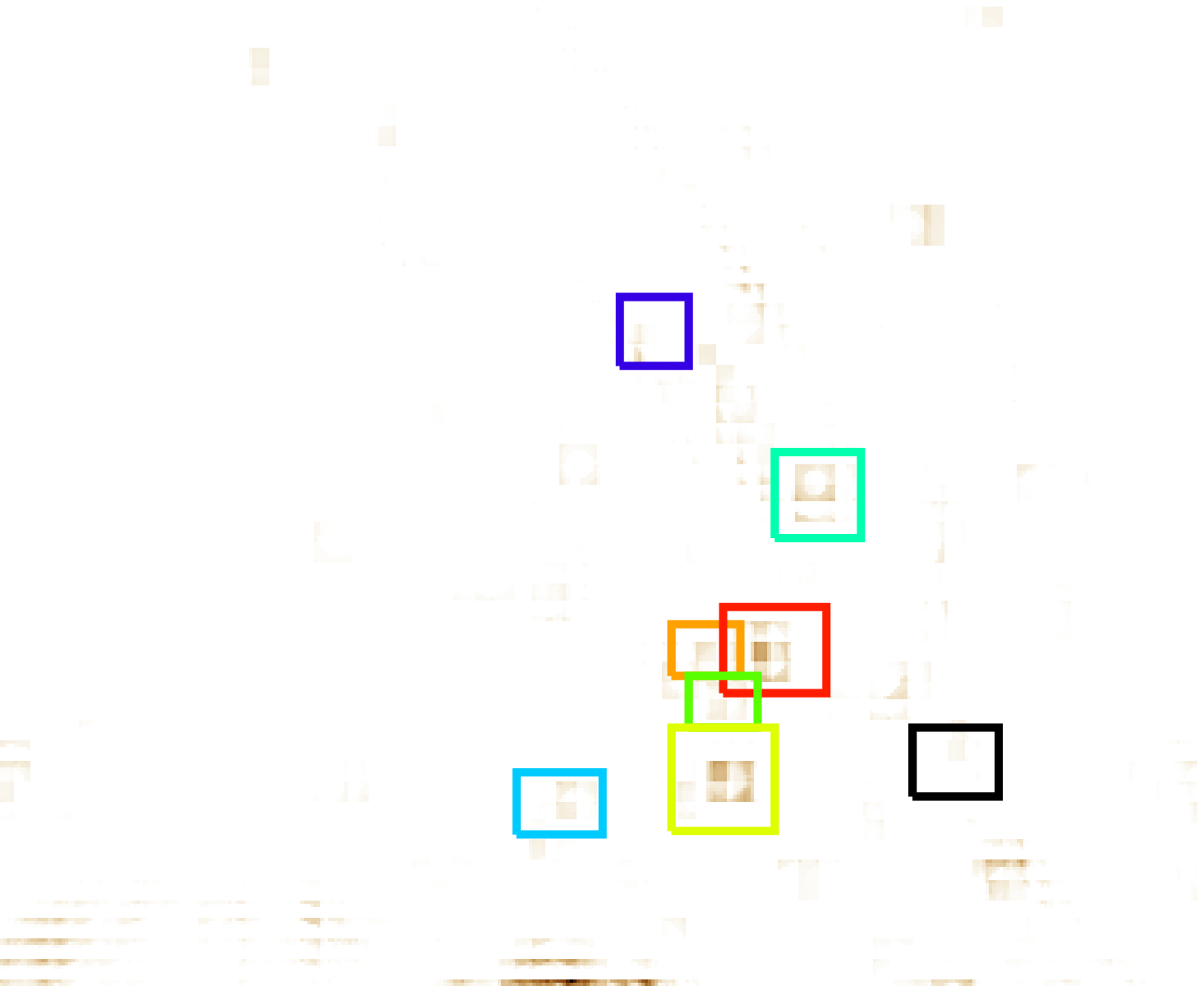}
\caption{Top row: integrated intensities over the whole Hi-C time series
observed in  Hi-C at 193 \AA\ (left), and in SDO/AIA at 193 \AA\ (right); Bottom
row: filtered and integrated intensities during the whole Hi-C time series:
(left) Hi-C EBDs, (right) SDO/AIA EBDs.}
\label{fig:hic_aia_int}
\end{center}
\end{figure}

In Fig.~\ref{fig:hic_aia_int}, we compare the EUV emission at 193\AA\ observed
in Hi-C (left) and SDO/AIA (right, \citeauthor{lemen12} \citeyear{lemen12}).
Even if centered on the same wavelength, the Hi-C and SDO/AIA filters are
similar but not identical. The high spatial resolution of the Hi-C observations
provides further insights into the small-scale coronal structures/phenomena
observed as localized brightenings; EBDs that can be resolved in the Hi-C FOV
are seen as compact bright regions in the SDO/AIA FOV. To extract the signature
of the EBDs, we proceed in three consecutive steps: (i) applying a median filter
to a single frame, (ii) considering the intensities above the 3$\sigma$-level
(standard deviation of the intensity map in a single frame), and (iii) adding
the intensities of the frames together over the whole Hi-C time series. We thus
can define the locations of significant bright dots (see
Fig.~\ref{fig:hic_aia_int} bottom left), and subsequently study their time
evolution and geometrical properties. Here, eight significant bright dots are
identified (as depicted by the rectangles in Fig.~\ref{fig:hic_aia_int} bottom
left). The eight different EBDs have been chosen as they are associated with a
well-defined structure in individual images: elliptical structures of more than
9 pixels and with  an intensity above the 3$\sigma$-level (standard deviation of
a single frame). Even if they do not obey the criteria imposed above, other
brightenings (not classified as EBDs) exist in the FOV, which suggest that
smaller spatial-scale structures can exist.

\begin{figure*}[!ht]
\begin{center}
\includegraphics[width=.16\linewidth, bb= 0 0 420 360]{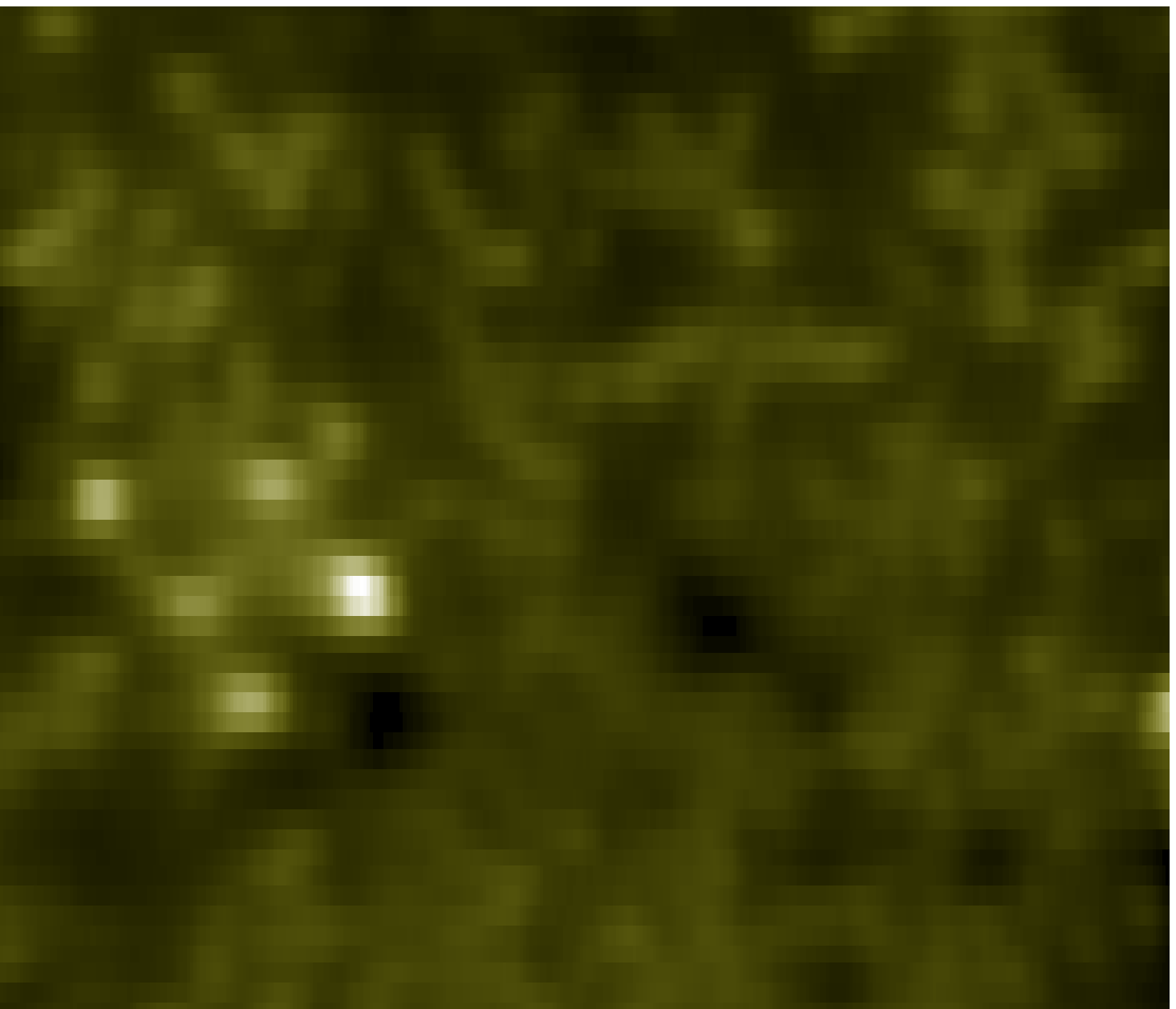}
\includegraphics[width=.16\linewidth, bb= 0 0 420 360]{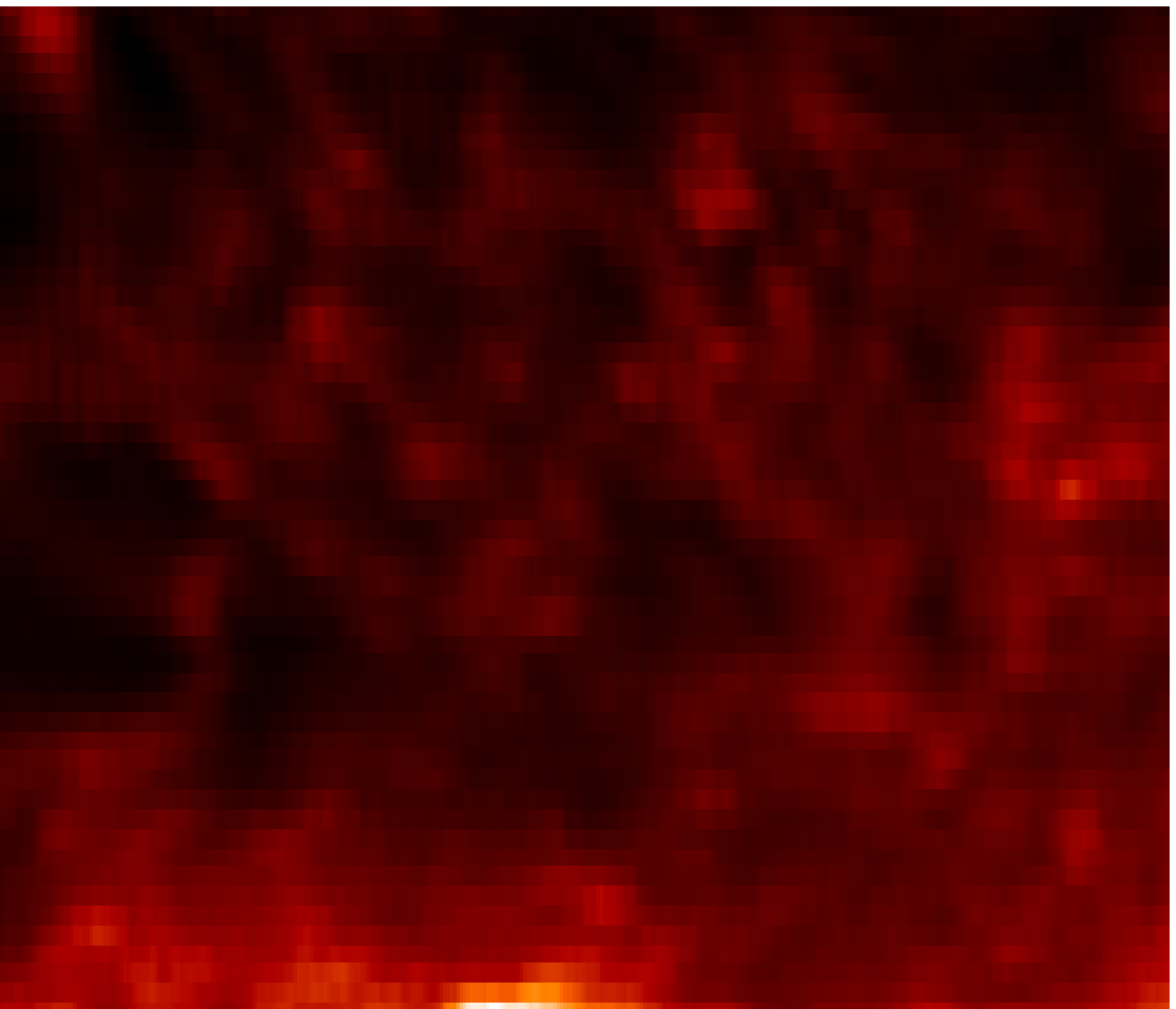}
\includegraphics[width=.16\linewidth, bb= 0 0 420 360]{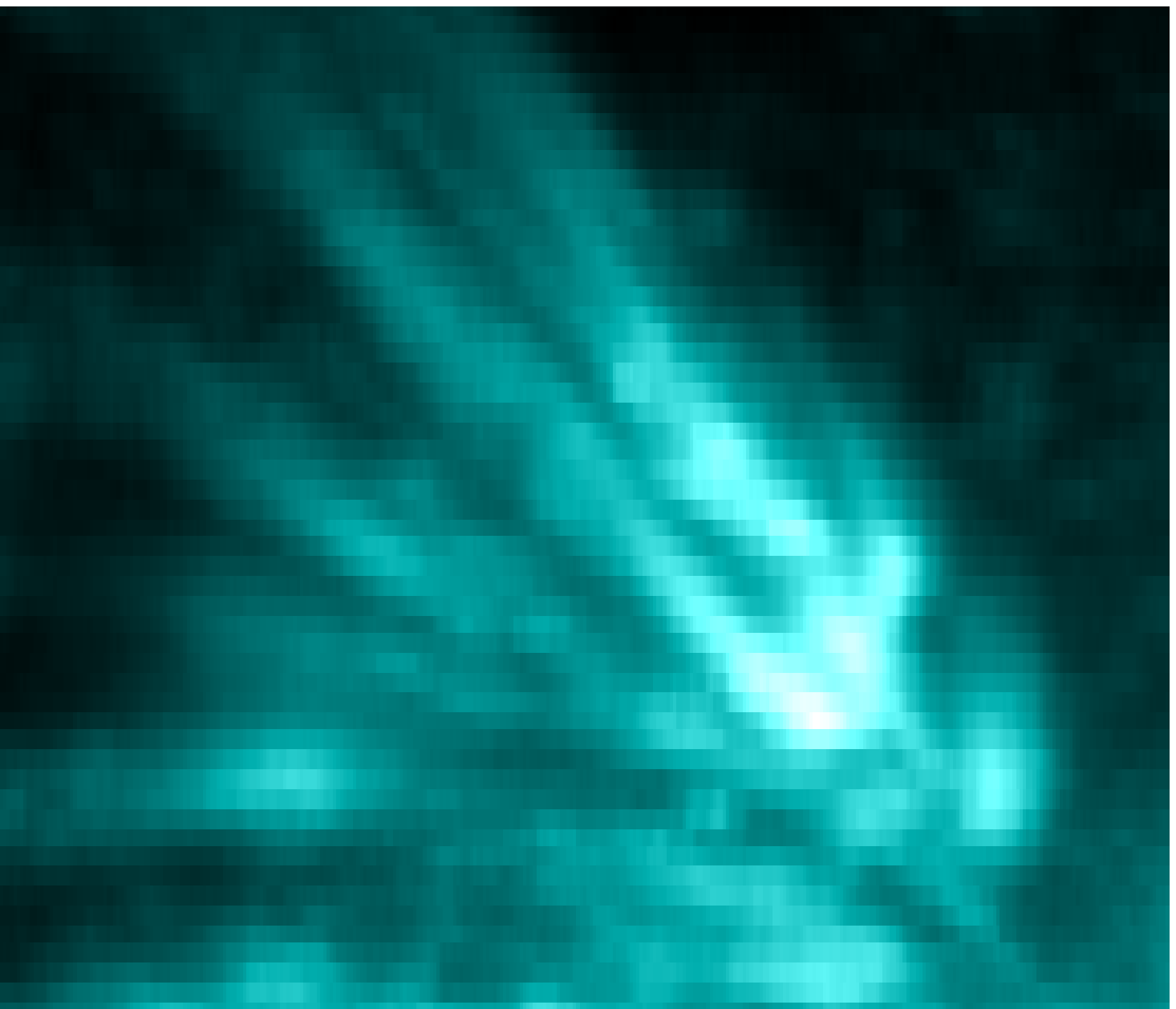}
\includegraphics[width=.16\linewidth, bb= 0 0 420 360]{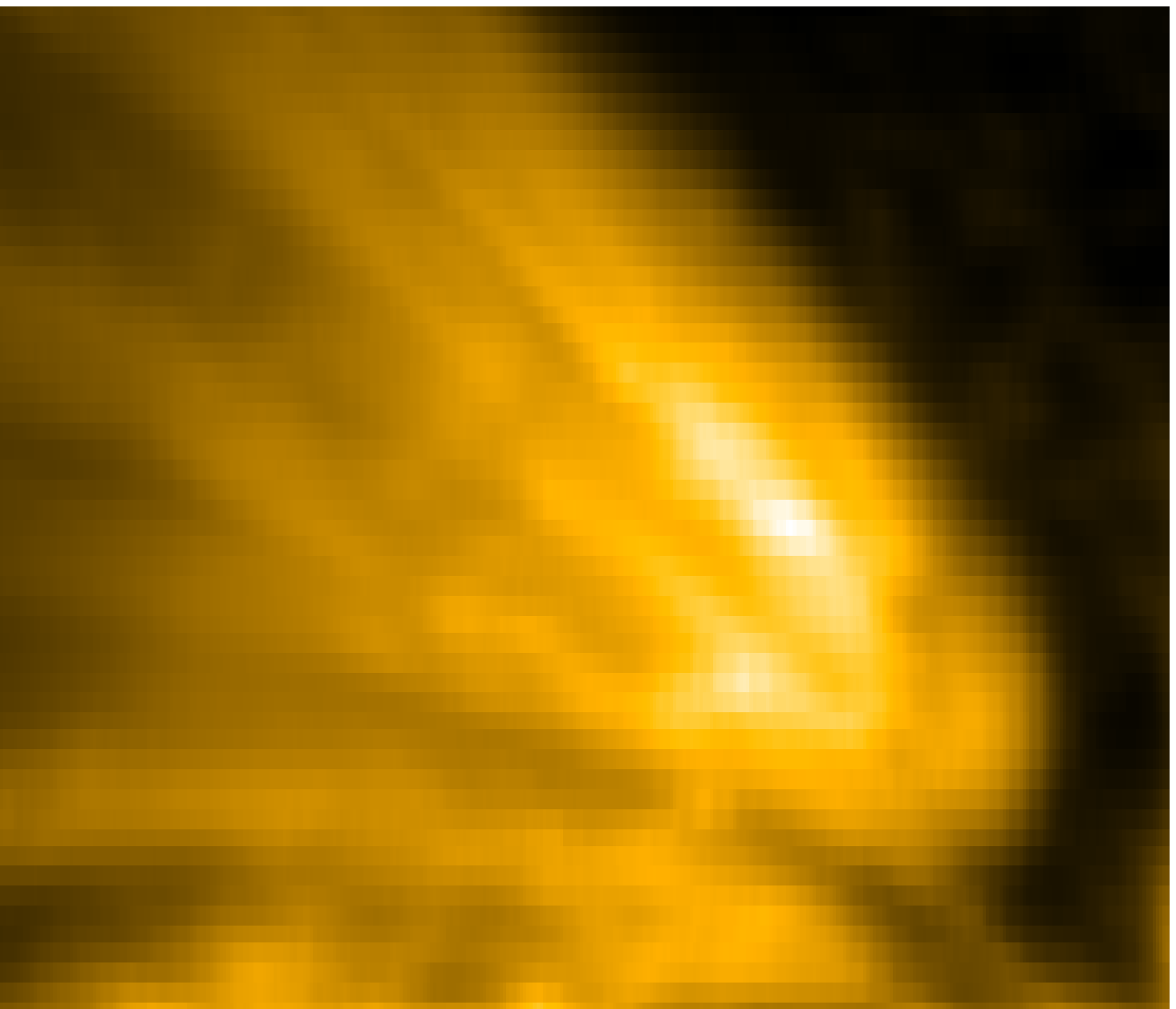}
\includegraphics[width=.16\linewidth, bb= 0 0 420 360]{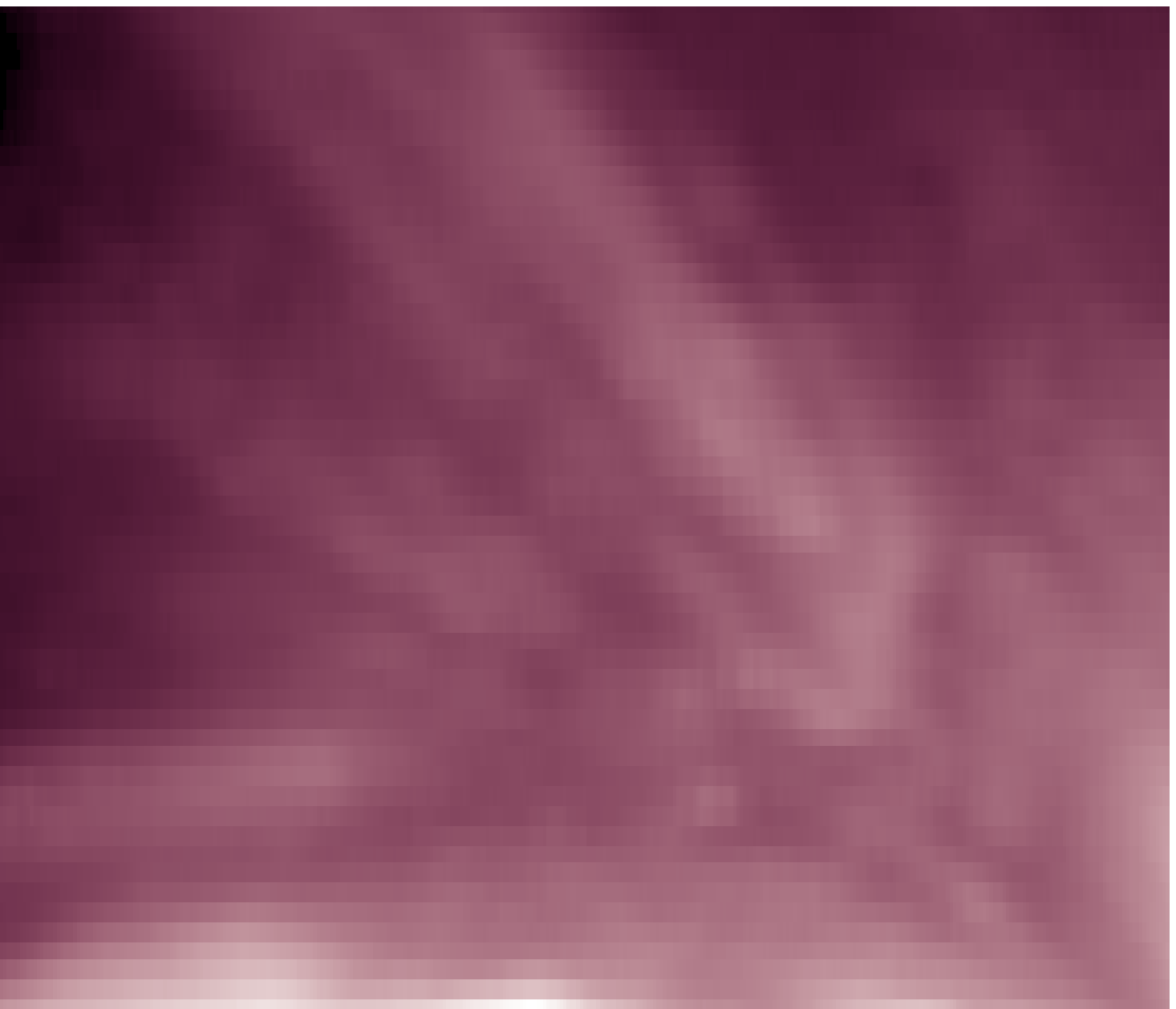}
\includegraphics[width=.16\linewidth, bb= 0 0 420 360]{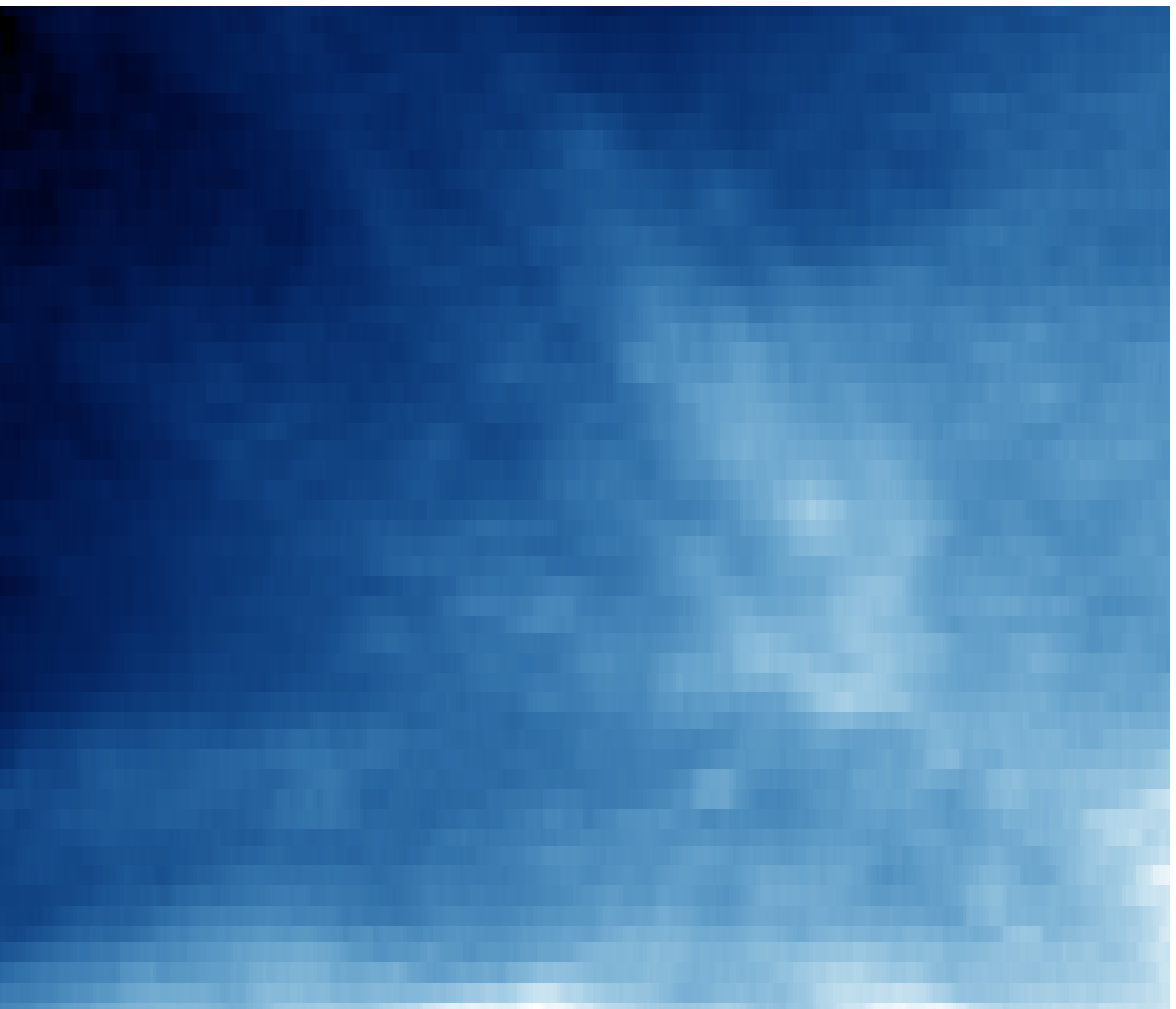}
\includegraphics[width=.16\linewidth, bb= 0 0 420 360]{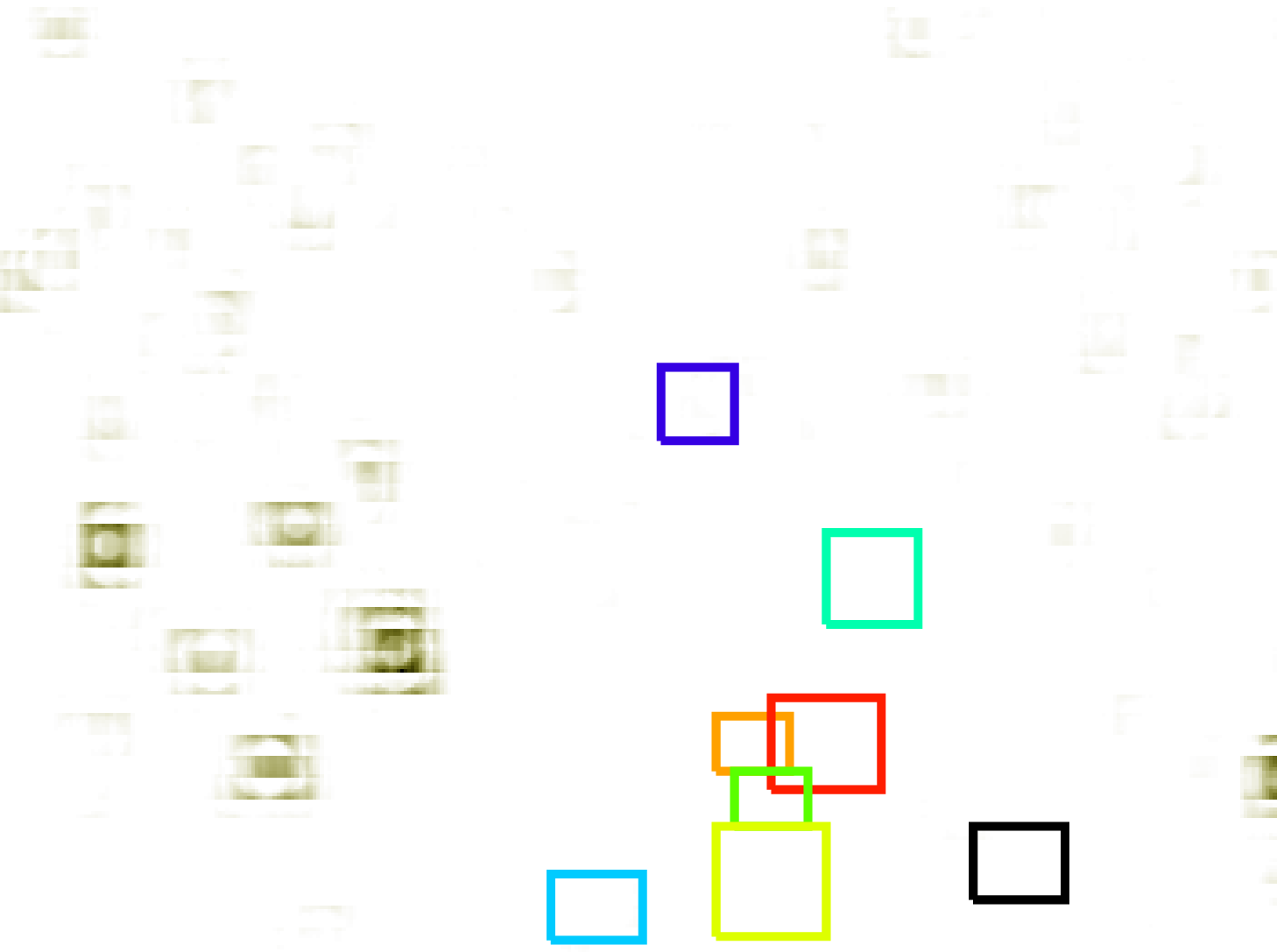}
\includegraphics[width=.16\linewidth, bb= 0 0 420 360]{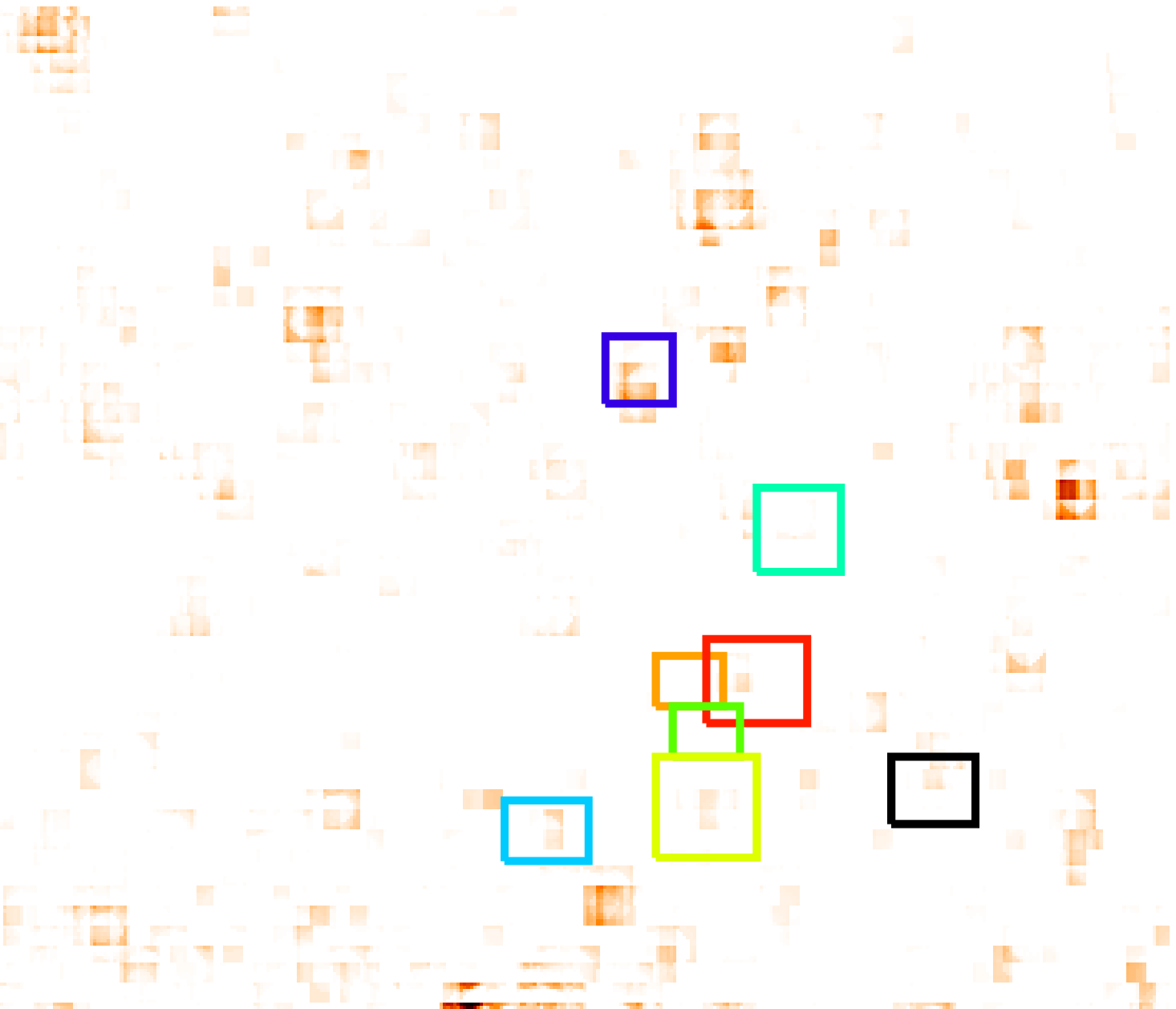}
\includegraphics[width=.16\linewidth, bb= 0 0 420 360]{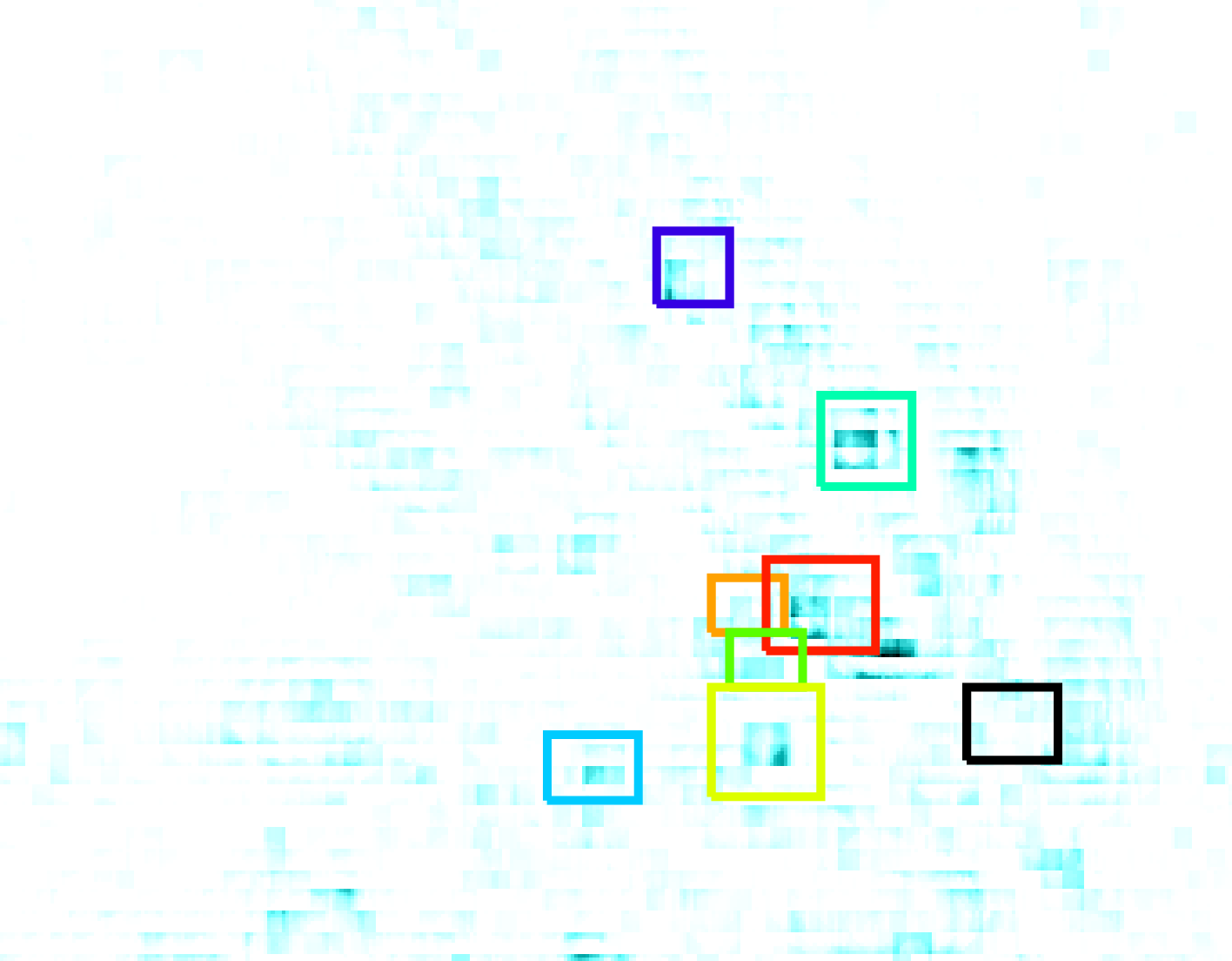}
\includegraphics[width=.16\linewidth, bb= 0 0 420 360]{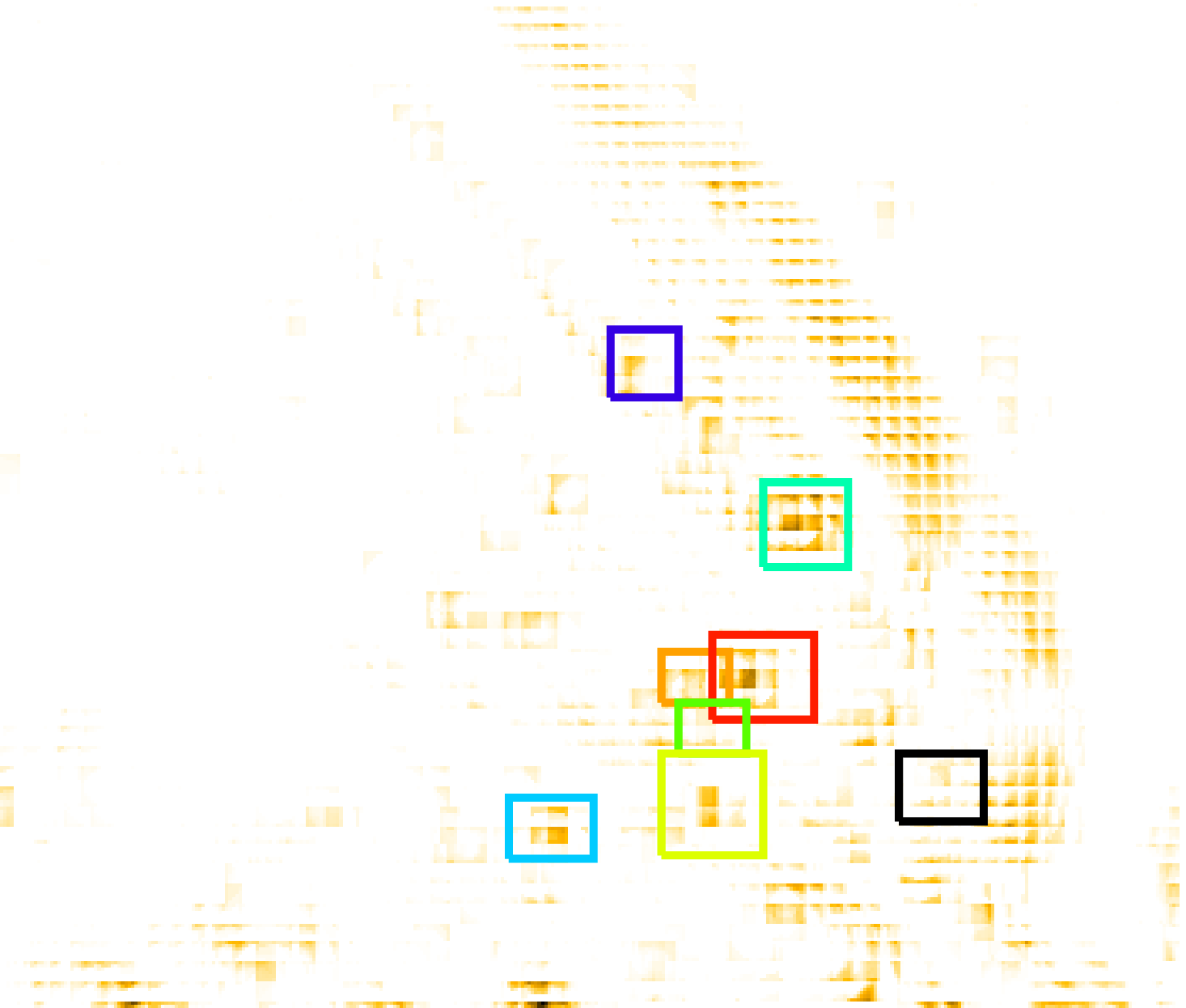}
\includegraphics[width=.16\linewidth, bb= 0 0 420 360]{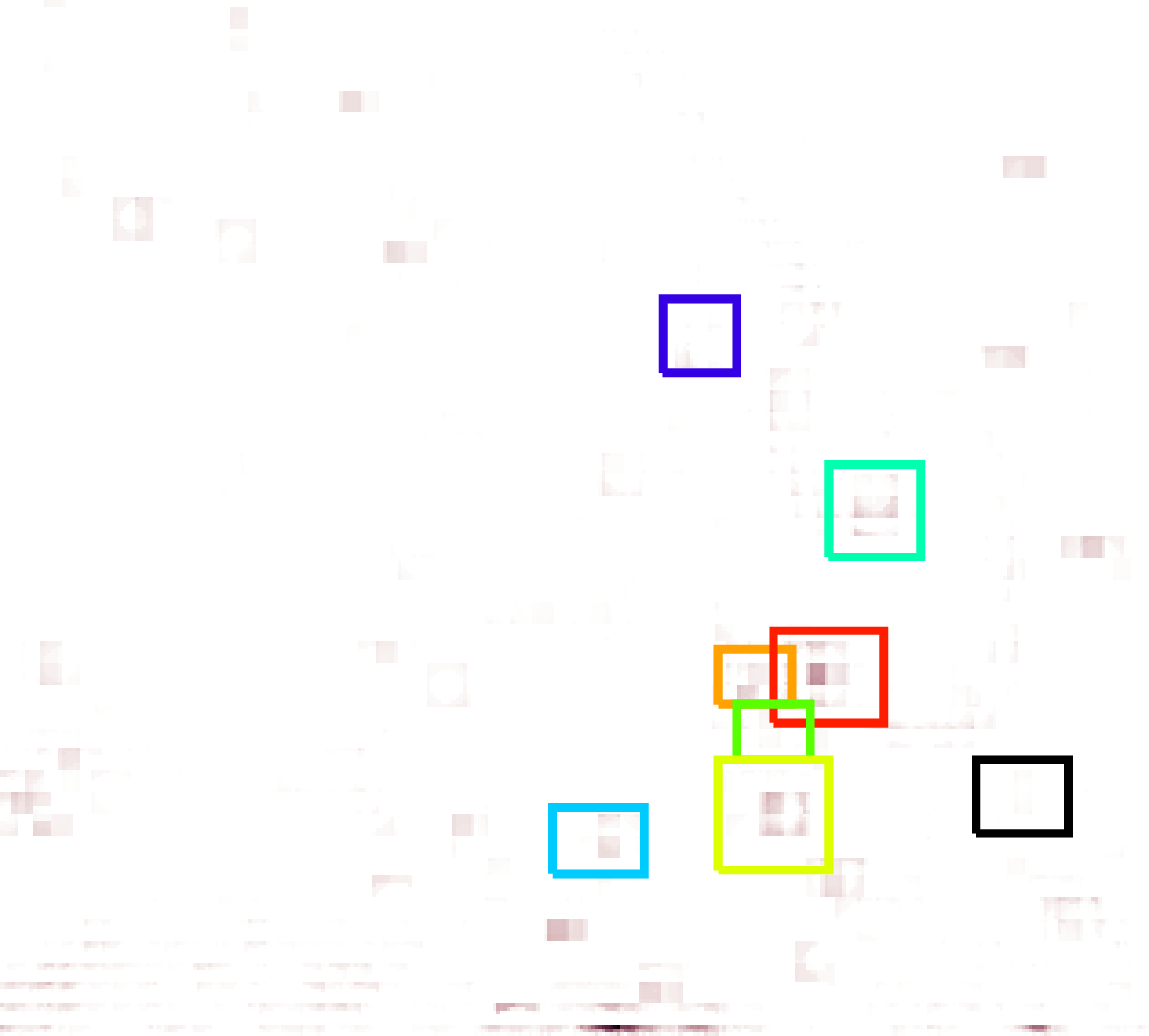}
\includegraphics[width=.16\linewidth, bb= 0 0 420 360]{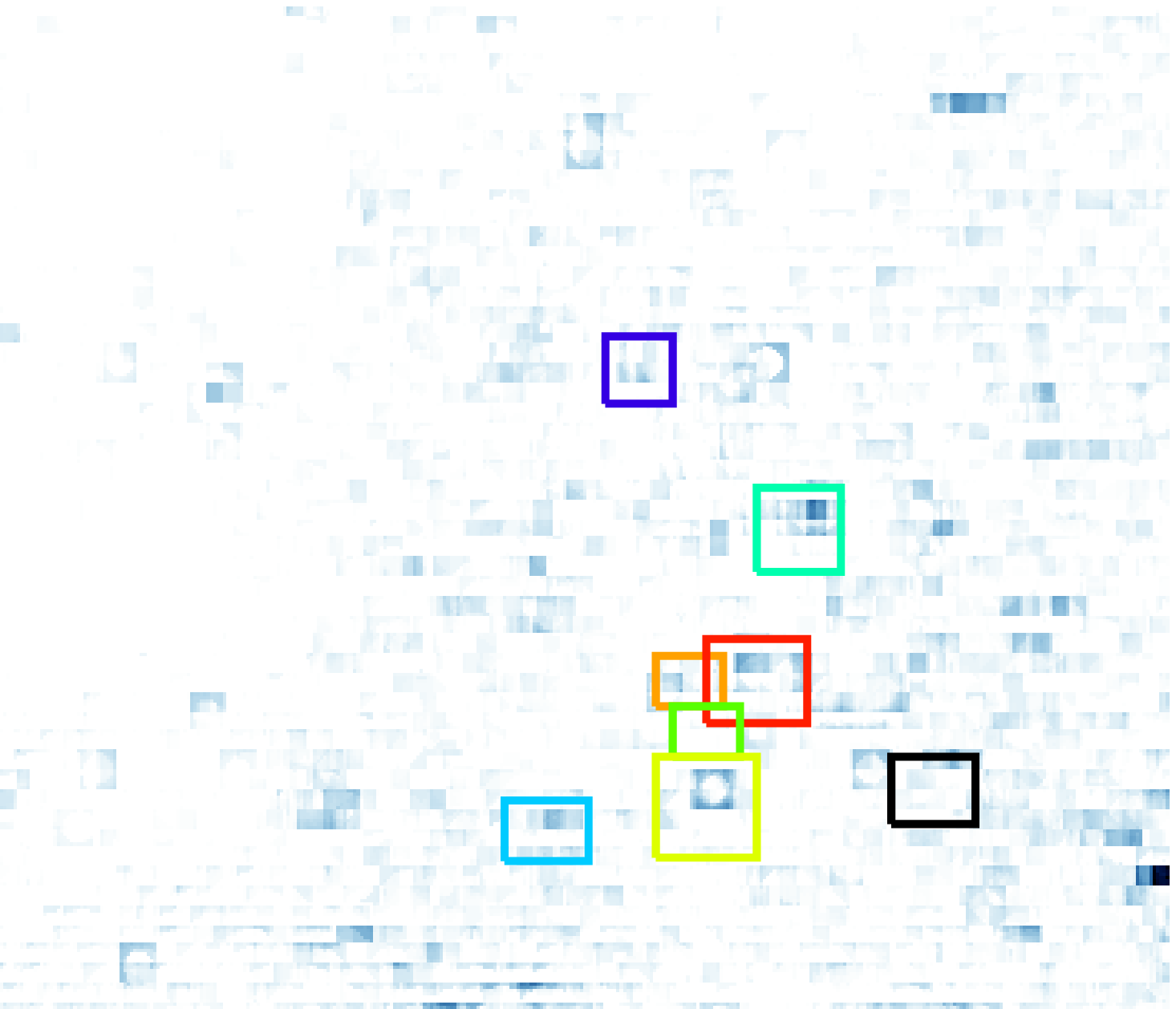}
\caption{(Top) Integrated intensity SDO/AIA images at 1600 \AA, 304 \AA, 131 \AA, 171 \AA,
211 \AA, 335 \AA\ (from left to right) in the same FOV and at the nearest time
to the Hi-C image in Fig.~\ref{fig:hic_aia_int}. (Bottom) filtered images as in
Fig.~\ref{fig:hic_aia_int} bottom (some color coding as in
Fig.~\ref{fig:hic_aia_int}). }
\label{fig:aia_im}
\end{center}
\end{figure*}

In Fig.~\ref{fig:hic_aia_int} bottom right, the location of the EBDs selected in
the Hi-C image (left) is compared to the SDO/AIA image (right) processed with
the same methodology as the Hi-C data: most of the chosen EBDs can be identified
in both fields-of-view. Solely based on the images, the signal is much weaker
for the SDO/AIA observations, and thus using SDO/AIA observations alone  would
not provide confidence of their existence; only the comparison with the Hi-C
images  ensures that the SDO/AIA EBDs are true brightenings and not just noise.

	\subsection{Multi-wavelength Observations of EBDs}

Using the capabilities of the SDO/AIA instrument, we study the EBDs using six
other wavelength channels covering from the low chromosphere to the high
temperature corona (1600\AA, 304\AA, 131\AA, 171\AA, 211\AA, and 335\AA). In
Fig.~\ref{fig:aia_im} top, we plot the integrated images over the whole Hi-C
time series in the different channels for the same FOV. In Fig.~\ref{fig:aia_im}
bottom, we apply the same data analysis procedure to the time series for the six
wavelength channels in order to enhance the location of EBDs.

From Fig.~\ref{fig:aia_im}, we find the following possible signatures of EBDs:
\begin{itemize}
\item[-]{the 1600\AA\ channel does not show much evidence of EBD intensity
signature.}
\item[-]{the 304\AA\ channel shows a weak emission for some of the EBDs.}
\item[-]{the hotter channels (131\AA, 171\AA, 211\AA, and 335\AA) show a clear
evidence of EBD intensity signatures as in the 193\AA\ channel.}
\end{itemize}

Comparing the temperature response functions of SDO/AIA and their overlapping
areas \citep{boe12}, this multi-channel analysis suggests that the EBDs are in a
temperature range between 0.3 to 1.5 MK ($\log(T) = 5.5-6.2$) which implies that
the EBDs are located at the top of the chromosphere or at the bottom of the
corona.

\begin{figure}[!h]
\begin{center}
\includegraphics[width=1.\linewidth]{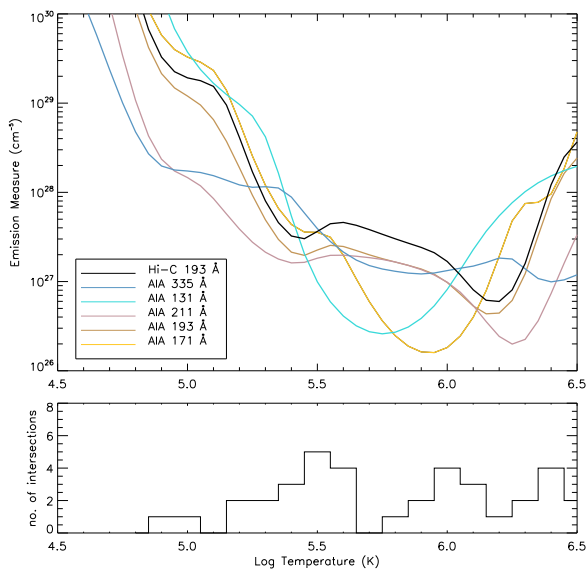}
\caption{EM loci for EBD3 at the peak of intensity. (Top) EM loci curves for the
different SDO/AIA channels and Hi-C intensity. (Bottom) Number of crossing at a
given temperature bin.}
\label{fig:eml_ebd3}
\end{center}
\end{figure}
	
	\subsection{EBD's Temperature: EM Loci} \label{sec:em_loci}

In order to investigate the possible temperature of the EBDs, the EM loci
technique was employed: the same method as in \citet{ale13} was used,
especially to get the temperature response functions of the different SDO/AIA
and Hi-C channels. The
EM loci method \cite[e. g.,][]{jor87, del02} allows us to determine if a coronal
structure is isothermal or multi-thermal without providing more in-depth information
(see Fig.~7 in \citeauthor{reale10} \citeyear{reale10} for an example of
isothermal and multi-thermal loops). We derived the EM loci
curves of several SDO/AIA channels (131\AA, 171\AA, 193\AA, 211\AA, 335\AA) and
the 193\AA\ Hi-C channel. A single crossing of all the EM loci curves should
indicated an isothermal plasma at this temperature.

In Fig.~\ref{fig:eml_ebd3}, we present the results of the EM loci method for
EBD3: there is a clear evidence that three possible temperature regions
corresponding to the larger number of loci crossings may be distinguished: one
at low temperature at $log(T) = 5.5$, and two at coronal temperature with
$log(T) = 6.0$ and $log(T) = 6.4$. However, for all the other EBDs studied, the
EM loci method does not lead to any conclusive temperature, the crossings being
distributed between $log(T) = 5.3-6.5$. The lack of clarity in the EM loci
curves maybe the result of two points:  either (i) the size and timing of the
EBD's structure at the SDO/AIA scale do not permit to draw any firm
conclusions (as noted in Section~\ref{sec:obs}, the EBDs are very difficult to
observe in SDO/AIA data), or (ii) the plasma of the EBDs is a transition region
plasma for which a temperature cannot be determined by the lines observed. The
latter will imply that the mechanism to release magnetic energy most likely
occurs at the footpoints of the coronal loops. The behavior of the EM loci
method for the EBDs as observed by SDO/AIA is consistent with the study of
multi-thermal plasma performed by \citet{gue12}.

\section{Characteristics and Dynamics of EBDs}
\label{sec:dyn}

	\subsection{EBD's Categories} \label{sec:ts_ebd}

\begin{table}[!h]
\begin{center}
\caption{Characteristic parameters of EBDs}
\label{tab:bright}
\begin{tabular}{cccccccc}
\tableline\tableline \\[-0.2cm]
& Cat. & Name & Peak time & $\Delta$I & \multicolumn{2}{c}{Duration (s)} & Length \\
 &  & & $t_{peak}$ (UT) & & $\Delta_{peak}$ & $\Delta_{v}$ & 
 	(Mm) \\[0.1cm]
\tableline \\[-0.2cm]
 & I & EBD1 & 18:52:48 & 1.68 & 27  & 64  & 0.65 \\[0.1cm]
 & & EBD2 & 18:52:59 & 2.00 & 39  & 61  & 0.5 \\[0.1cm]
 & & EBD3 & 18:54:06 & 3.89 & 22  & 89  & 0.86 \\[0.3cm]
\rotatebox{90}{\makebox(0,0)[lc]{~\parbox{2cm}{\quad Single}~}}& 
\multicolumn{3}{l}{Average values} & 2.52 & 29s & 71s 
	& 0.67 Mm \\[0.1cm]
\hline \\[-0.2cm]
& II & EBD4 & 18:52:59 & 1.99 & 28  & 61  & 0.43 \\[0.1cm]
 &  & & 18:53:16 & 1.82 & 22  &     & 0.76 \\[0.1cm]
 & & EBD5 & 18:53:10 & 1.55 & 23  & 56  & 0.76 \\[0.1cm]
  & & & 18:53:43 & 1.52 & 28  &     & 0.72 \\[0.1cm]
& & EBD6 & 18:54:56 & 1.76 & 16  & 83  & 0.65 \\[0.1cm]
  & &  & 18:55:12 & 1.84 & 22  &     & 0.76 \\[0.3cm]
\rotatebox{90}{\makebox(0,0)[lc]{~\parbox{4cm}{\quad \quad Double-peak}~}}& 
\multicolumn{3}{l}{Average values} & 1.75 & 23s & 67s 
	& 0.68 Mm \\[0.1cm]
\hline \\[-0.2cm]
& III & EBD7 & 18:52:31 & 1.34 & 23 & 200 & D \\[0.1cm]
 & & & 18:52:59 & 1.70 & 23 &  & 0.43 \\[0.1cm]
 & & & 18:53:43 & 1.98 & 28 & & 0.79 \\[0.1cm]
 & & & 18:54:17 & 2.06 & 23 & & 0.79 \\[0.1cm]
 & & & 18:54:45 & 1.93 & 5 & & 0.57 \\[0.1cm]
 & & & 18:55:01 & 2.38 & 16 & & 0.79 \\[0.1cm]
\rotatebox{90}{\makebox(0,0)[lc]{~\parbox{4cm}{\quad Long Duration}~}}& 
\multicolumn{3}{l}{Average values} & 1.89 & 20s & 200s 
	& 0.67 Mm \\[0.1cm]
\hline \\[-0.2cm]
& IV & EBD8 & 18:53:21 & 1.28 & 5 & 133 & 0.43 \\[0.1cm]
 & & & 18:53:43 & 1.32 & 5 & & 0.43 \\[0.1cm]
 & & & 18:54:00 & 1.47 & 5 & & 0.43 \\[0.1cm]
 & & & 18:54:45 & 1.44 & 5 & & 0.43 \\[0.1cm]
 & & & 18:54:56 & 1.44 & 5 & & 0.43 \\[0.1cm]
\rotatebox{90}{\makebox(0,0)[lc]{~\parbox{2cm}{\quad \quad Bursty}~}}& 
\multicolumn{3}{l}{Average values} & 1.39 & 5s & 133s 
	& 0.43 Mm \\[0.1cm]
\tableline
\end{tabular}
\end{center}
\end{table}


We analyse the Hi-C light curves and images of the eight selected EBDs within
the areas depicted in Fig.~\ref{fig:hic_aia_int} bottom left.  In
Table~\ref{tab:bright}, we give the main observed characteristics for the eight
chosen EBDs including the time of the peak intensity $t_{peak}$, the relative
intensity $\Delta I$ of the maximum intensity at  $t_{peak}$ with respect to the
average intensity over the full EBD FOV at the same time, the duration
$\Delta_{peak}$ from the light curve given by the width at half-maximum of the
intensity peak (with respect to the minimum intensity of the light curve), and
$\Delta_{v}$ the time during which a coherent structure (elliptical surface of
intensity covering more than 9 pixels) can be identified within the time series,
and the characteristic length (major axis of the ellipse) of the EBDs. The area
considered for individual EBDs is the size of the rectangle depicted in
Fig.~\ref{fig:hic_aia_int} bottom left. 

From Table~\ref{tab:bright}, we deduce that, in average, EBDs have a
characteristic life-time of 25s, a characteristic length of 0.68 Mm.

From the light curves of the identified EBDs, we distinguish four
categories of EBDs based on their frequency within the Hi-C time series (see
Figs~\ref{fig:lc_ebd3}--\ref{fig:lc_ebd8} top, and also Table~\ref{tab:bright}):
\begin{itemize}
\item[{\bf I.}]{{\bf Single Peak} single events with a clear peak in the light
curve. EBD1--3 have a duration between 22 and 39s, and a characteristic size
between 0.5 and 0.86 Mm;}
\item[{\bf II.}]{{\bf Double Peak} double-peaked EBDs with two consecutive
maxima. EBD4--6 have two peaks separated by 17--33s, and with a short global
duration (about 67s);}
\item[{\bf III.}]{{\bf Long Duration} long-duration EBDs for which the coherent
structure can be characterized during the entire time series (200s), and with
several distinct maxima. For EBD7, six intensity peaks are observed within the
Hi-C time series with a time interval from 16s to 44s. In
Table~\ref{tab:bright}, the value "D" means that the EBD intensity distribution
is diffuse;}
\item[{\bf IV.}]{{\bf Bursty} a series of very short brightenings related to a
coherent structure, amd leading to an increase of the overall intensity. EBD8
light curve shows an increase of intensity starting at 18:53:16 UT, and has
several short peaks (5s) observed with a cadence from 11s to 45s.}
\end{itemize} 

From this small sample of events, Categories I and II are the most
significant. We now proceed to a detailed analysis of one single EBD in each
of the four categories. For Categories I and II, the EBD analysed has the
largest intensity contrast ($\Delta$I in Table~\ref{tab:bright}).

	\subsection{Category I: Single Peak}

\begin{figure}[!ht]
\begin{center}
\includegraphics[width=.7\linewidth]{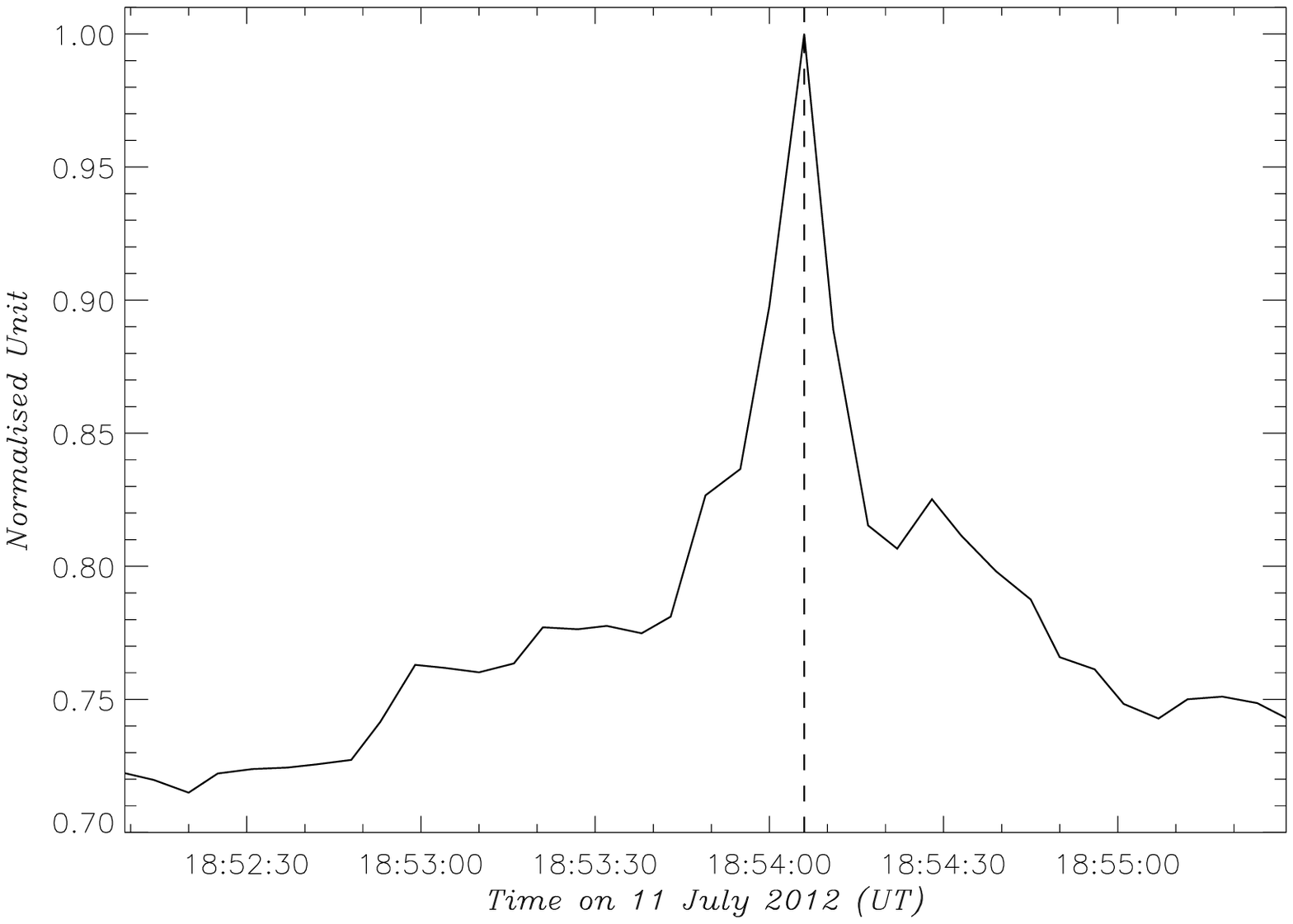}
\includegraphics[width=.7\linewidth]{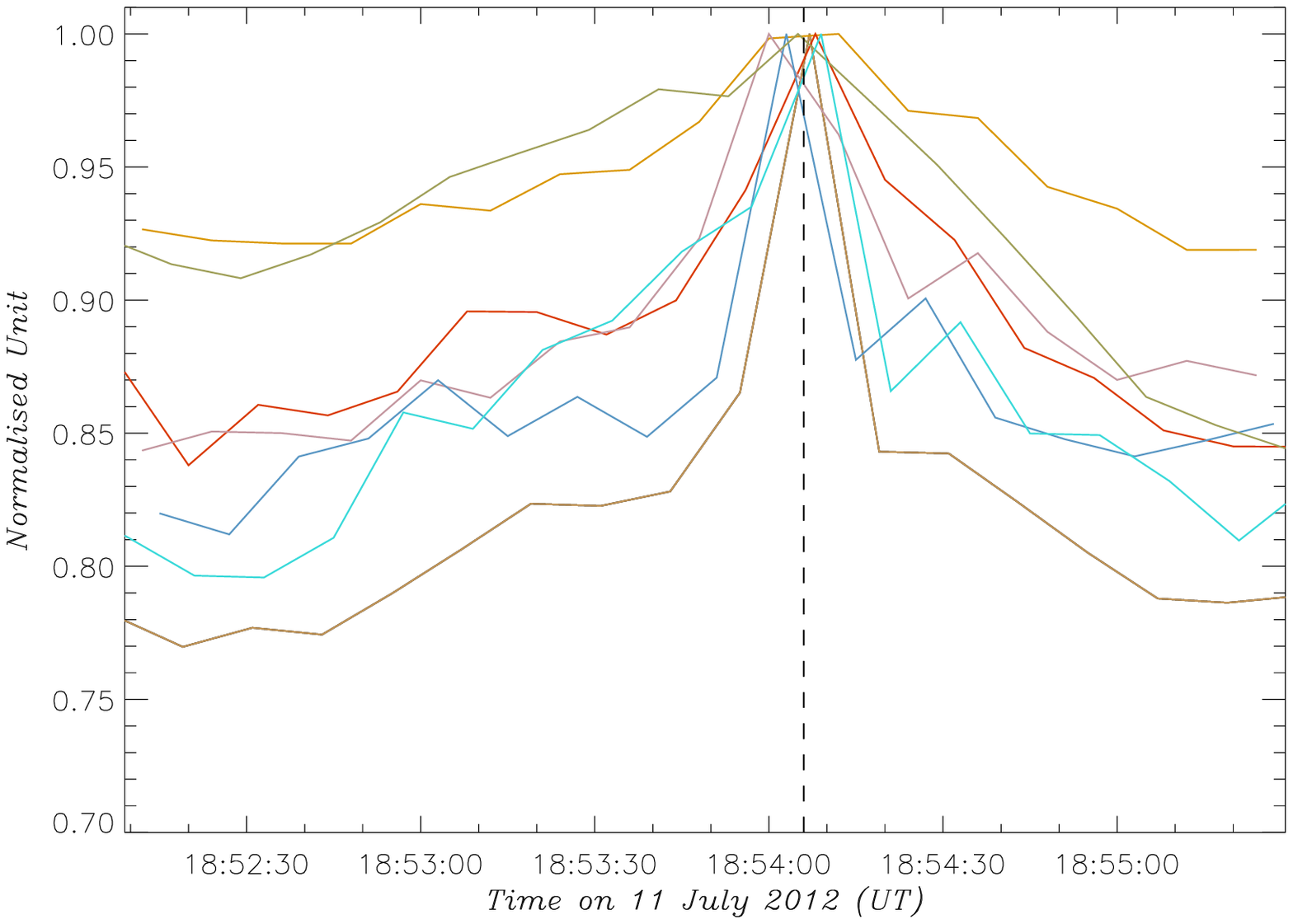}
\includegraphics[width=0.6\linewidth, bb = 0 50 504 150]{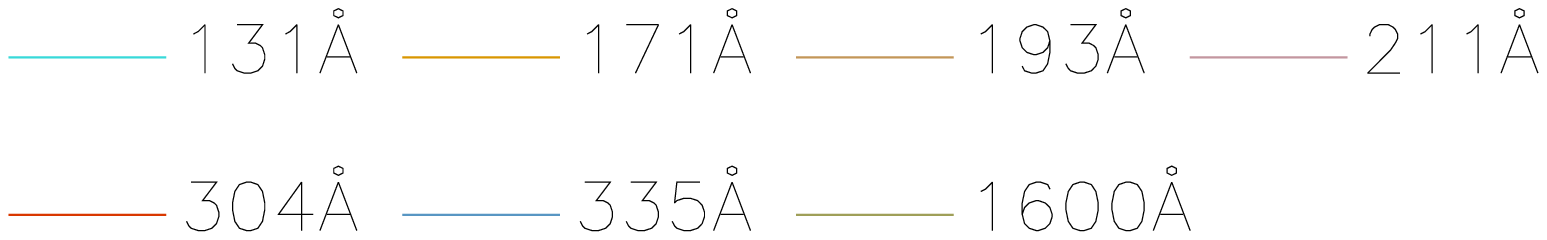}
\caption{(Top) Hi-C normalised light curve for EBD3 showing clearly the single
peak of intensity charaterising the bright dot. (Bottom) SDO/AIA normalised
light curves for the same area encompassing EBD3 in seven different EUV-UV
channels (131\AA, 171\AA, 193\AA, 211\AA, 304\AA, 335\AA, and 1600\AA)}
\label{fig:lc_ebd3}
\end{center}
\end{figure}
	
Consider EBD3; in Fig.~\ref{fig:lc_ebd3} top, we plot the Hi-C normalised light
curve of EBD3, which exhibits a clearly distinctive peak in intensity. The peak
lasts during 5 frames (22 s). We also plot the normalised light curves for
different SDO/AIA channels (see  Fig.~\ref{fig:lc_ebd3} bottom) where we
distinguish a corresponding peak in intensity (see Fig.~\ref{fig:ts_ebd3}a). Due
to the short duration of the EBDs ($\sim$22s) and the time cadence of SDO/AIA
images (12s), it is difficult to conclude when the peak emission has actually
occurred in the different channel. As shown by \cite{via11,via12}, the cooling
of a nanoflare storm within a coronal loop may be deduced from the ordering of
the appearance of intensity peaks in the SDO/AIA light curves. This result
follows the observational work done by \cite{win03} using TRACE data. Such
analysis cannot be performed here.

\subsection{Category II: Double Peak}

\begin{figure}[!ht]
\begin{center}
\includegraphics[width=.7\linewidth]{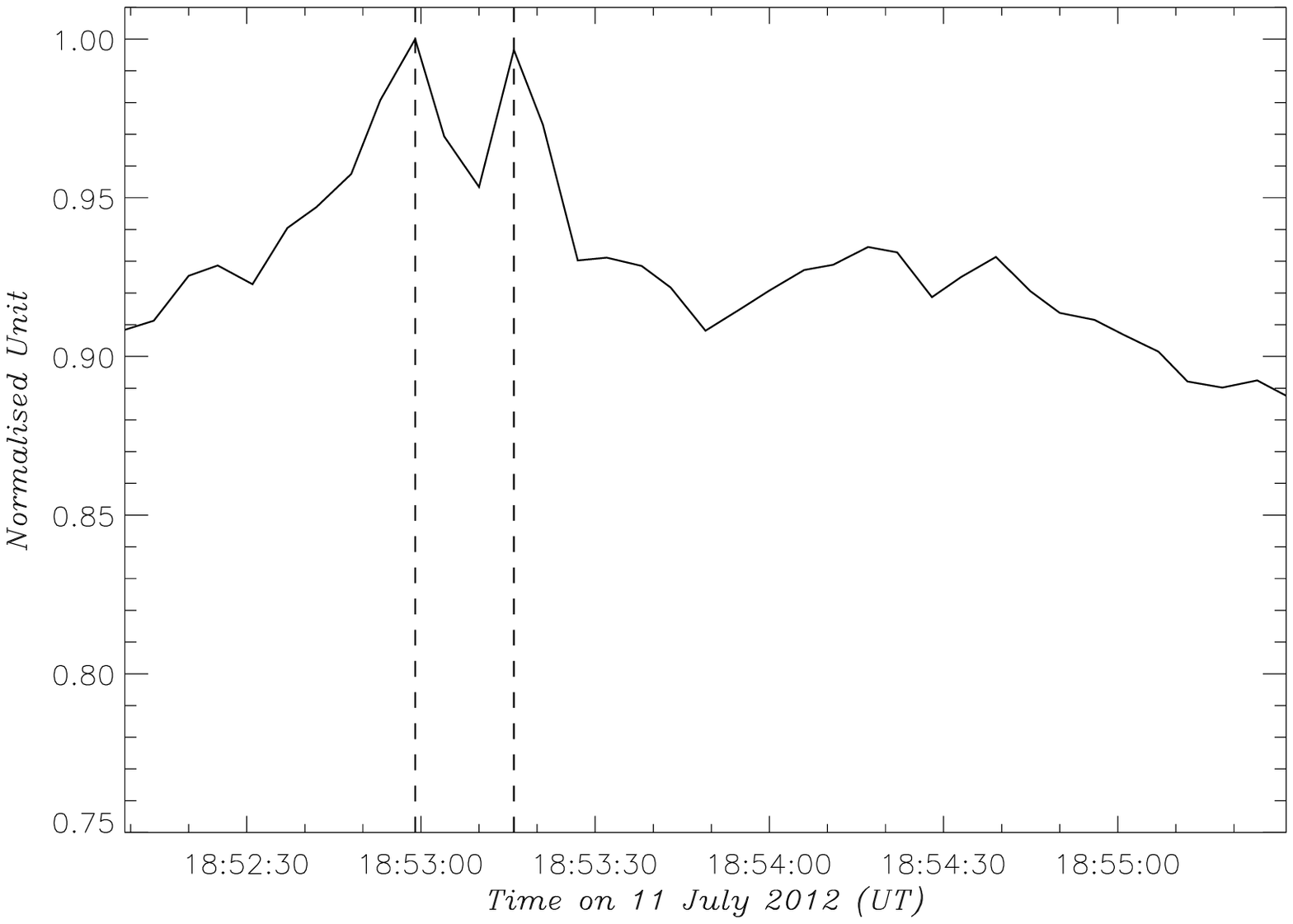}
\includegraphics[width=.7\linewidth]{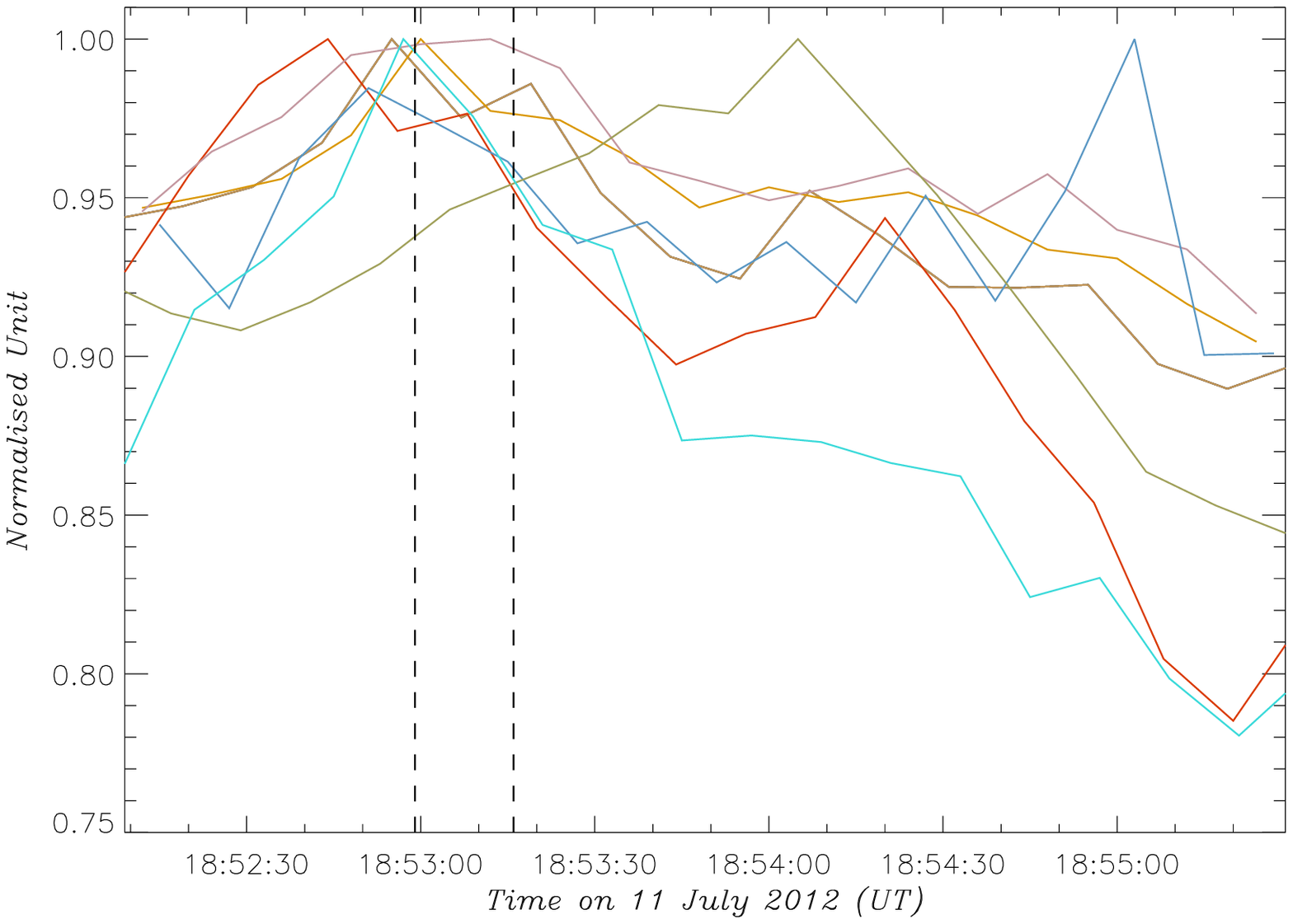}
\includegraphics[width=0.6\linewidth, bb = 0 50 504 150]{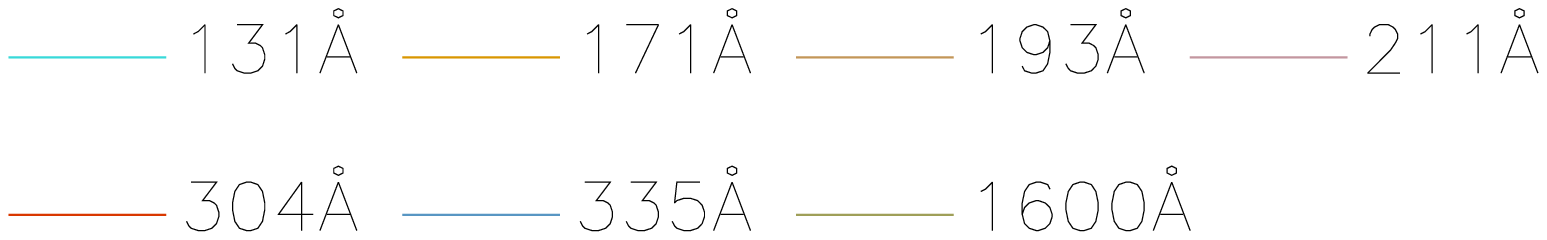}
\caption{Same as Fig.~\ref{fig:lc_ebd3} for EBD4 in Category II}
\label{fig:lc_ebd4}
\end{center}
\end{figure}

In Fig.~\ref{fig:lc_ebd4} top, the normalised Hi-C light curve for EBD4 clearly
shows two disctinct consecutive peaks (see Fig.~\ref{fig:ts_ebd3}b): each
individual peak has the property of the EBDs of Category I. In
Fig.~\ref{fig:lc_ebd4} bottom, the peaks are also observed in the 193\AA\ and
171\AA\ SDO/AIA channels. The first peak at 18:52:59 UT is clearly observed in
the other channels, while the second peak cannot be identified.  

\subsection{Category III: Long Duration}

\begin{figure}[!ht]
\begin{center}
\includegraphics[width=.7\linewidth]{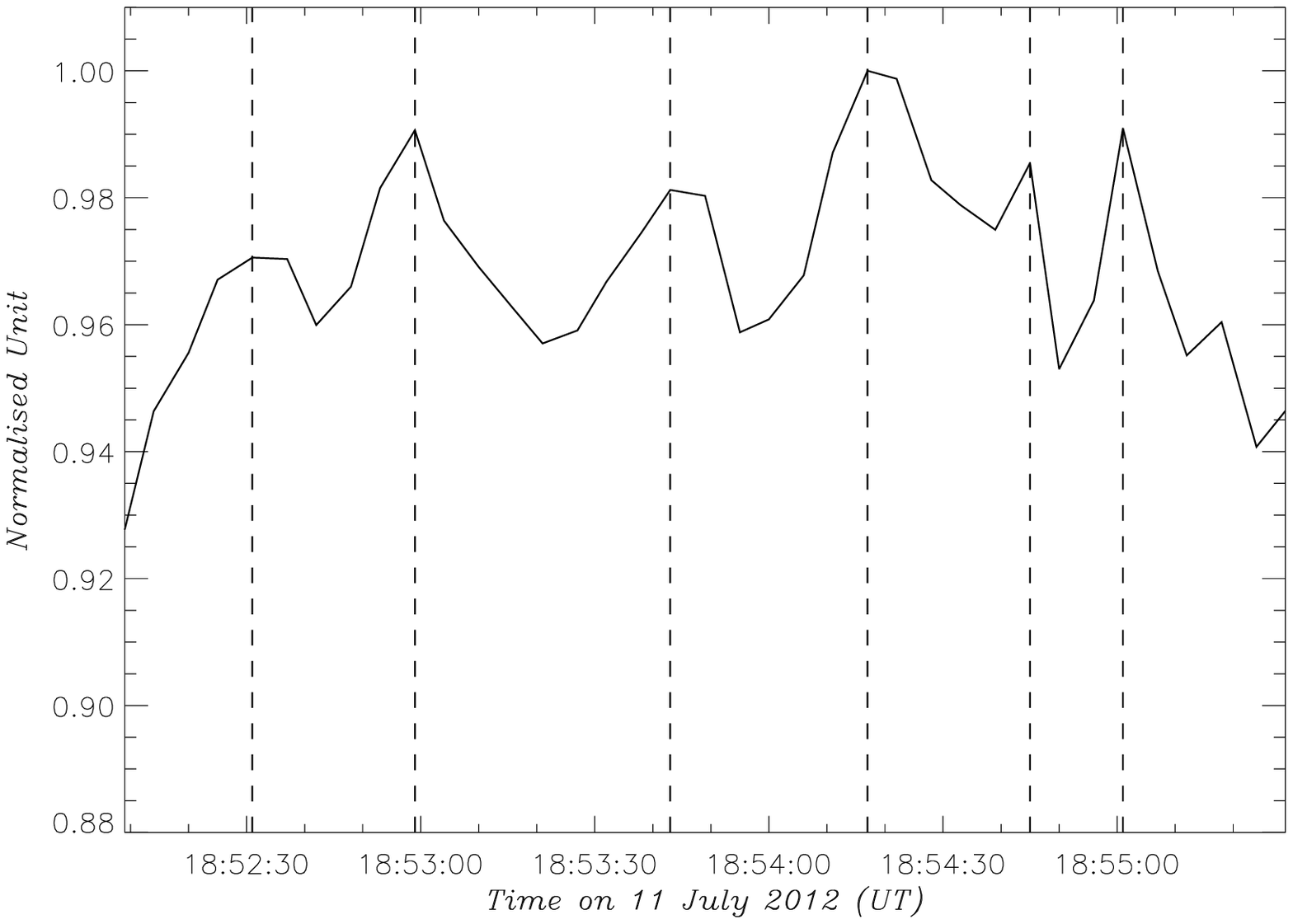}
\includegraphics[width=.7\linewidth]{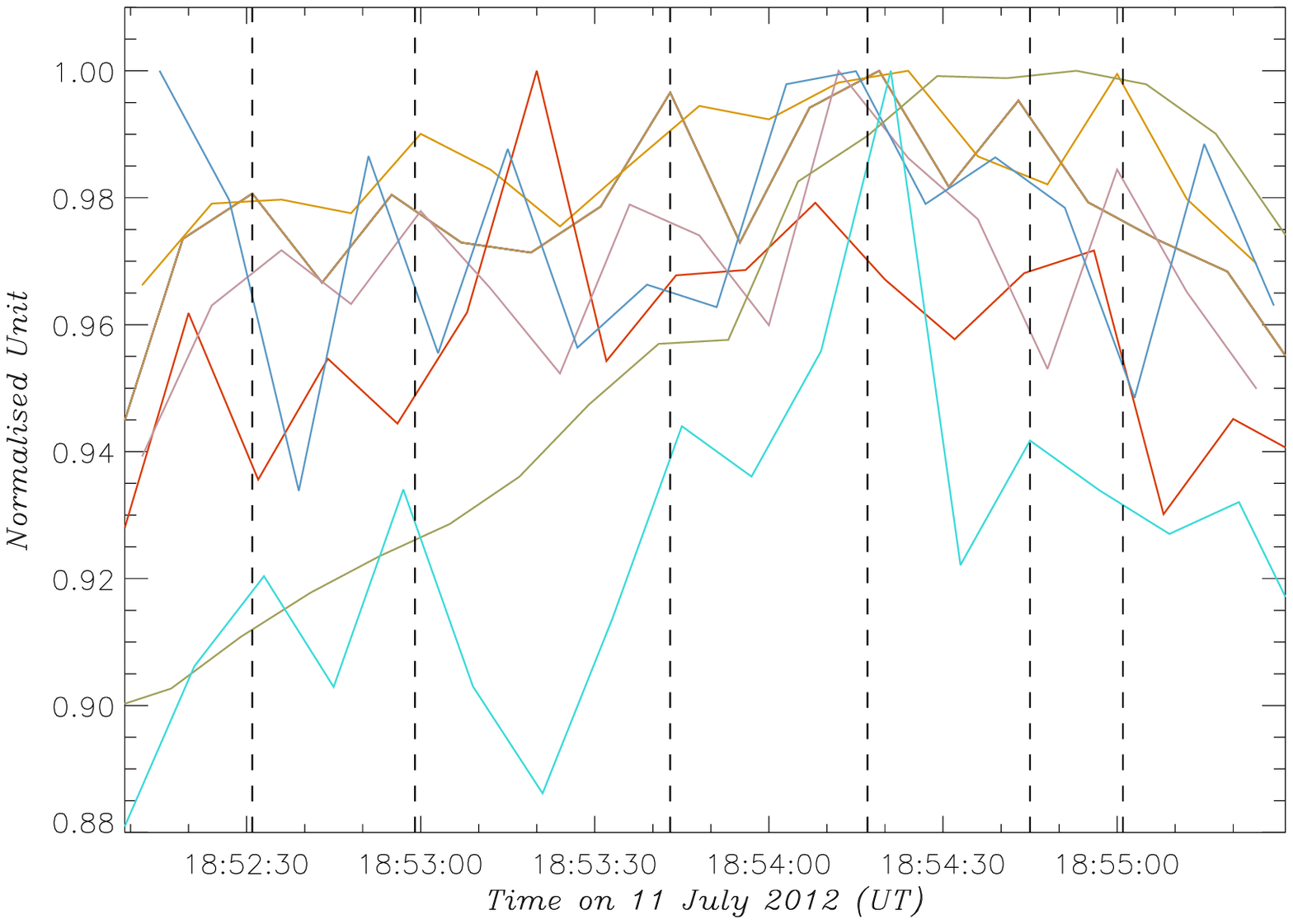}
\includegraphics[width=0.6\linewidth, bb = 0 50 504 150]{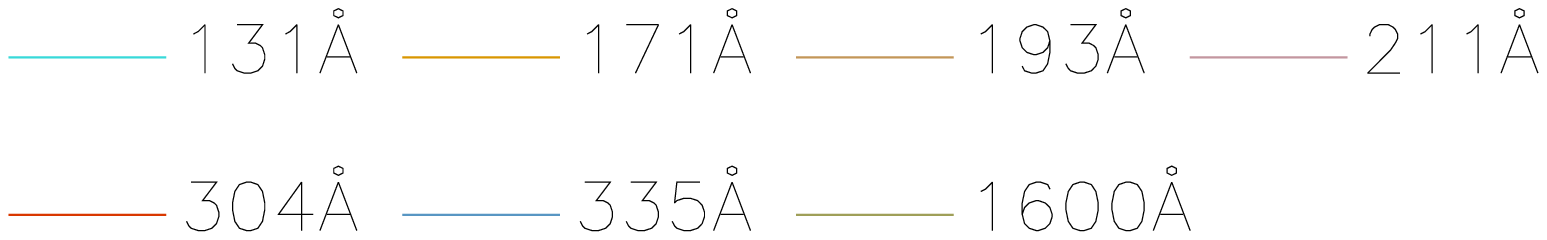}
\caption{Same as Fig.~\ref{fig:lc_ebd3} for EBD7 in Category III}
\label{fig:lc_ebd7}
\end{center}
\end{figure}


In Fig.~\ref{fig:lc_ebd7}, we plot the Hi-C normalised light curve (top) and the
SDO/AIA light curves (bottom) for EBD7, the only example of Category III EBDs.
The Hi-C light curve shows several intensity peaks that can be observed during
the whole time period. The peaks (marked by vertical dashed lines) appear at
18:52:31 UT, 18:52:59 UT, 18:53:43 UT, 18:54:17 UT, 18:54:45 UT, and 18:55:01
UT. These Hi-C peaks are also observed in the 193\AA\ SDO/AIA channel with a
comparable mean intensity value, and most of them can be seen in the 131\AA,
171\AA. We also notice that the normalised light curves of 211\AA, 304\AA\ and
335\AA\ exhibit a similar bursty behavior but at different time. EBD7 is
observed as a coherent structure during the whole Hi-C time series (see
Fig.~\ref{fig:ts_ebd3}c). It is also
possible to observe a slight displacement of the center-of-mass of EBD7, which
leads to an apparent transverse velocity of 4.2  km$\cdot$s$^{-1}$ for a
duration of 106s. The EBDs recur at a period of about 30s.

\subsection{Category IV: Bursty}

\begin{figure}[!ht]
\begin{center}
\includegraphics[width=.7\linewidth]{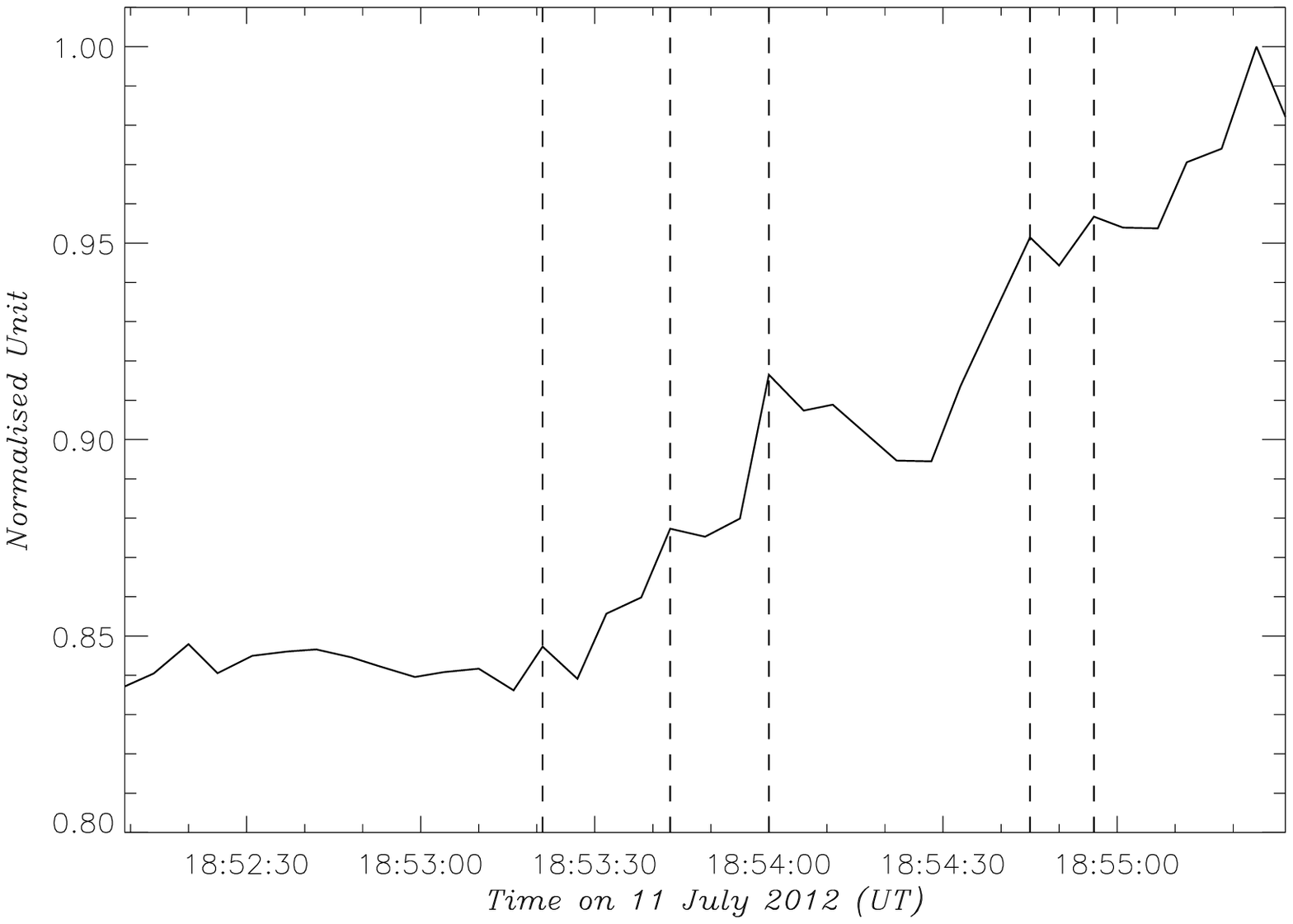}
\includegraphics[width=.7\linewidth]{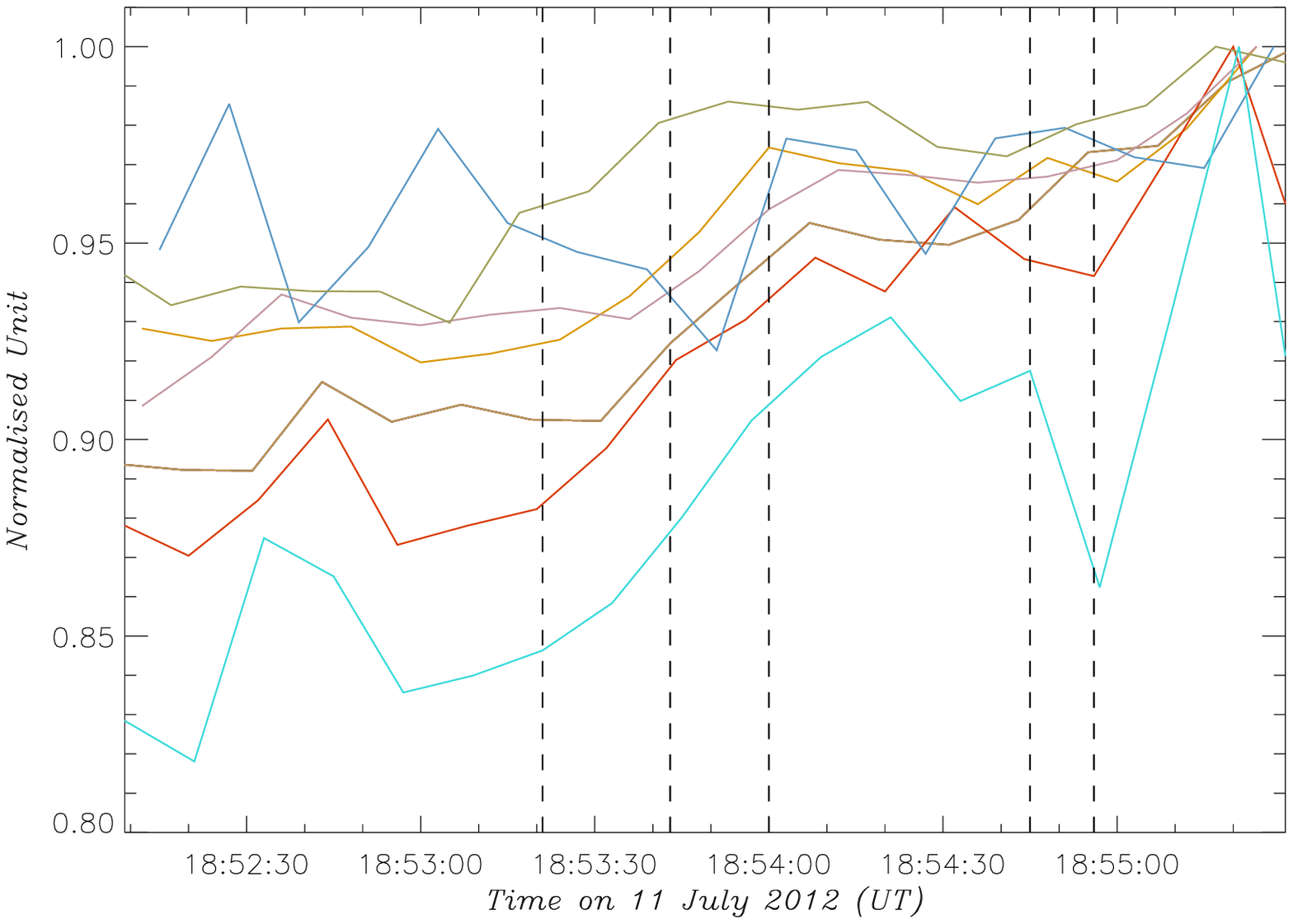}
\includegraphics[width=0.6\linewidth, bb = 0 50 504 150]{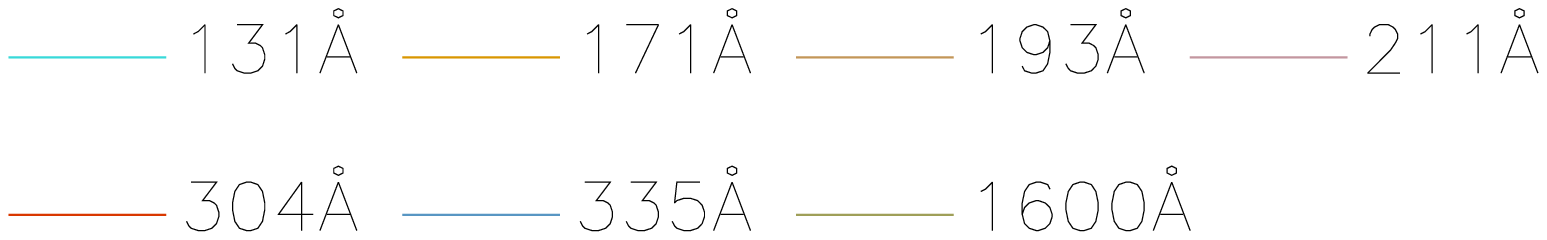}
\caption{Same as Fig.~\ref{fig:lc_ebd3} for EBD8 in Category IV}
\label{fig:lc_ebd8}
\end{center}
\end{figure}

\begin{figure*}[!t]
\begin{center}
\includegraphics[width=.257\linewidth, bb=0 0 430 360]{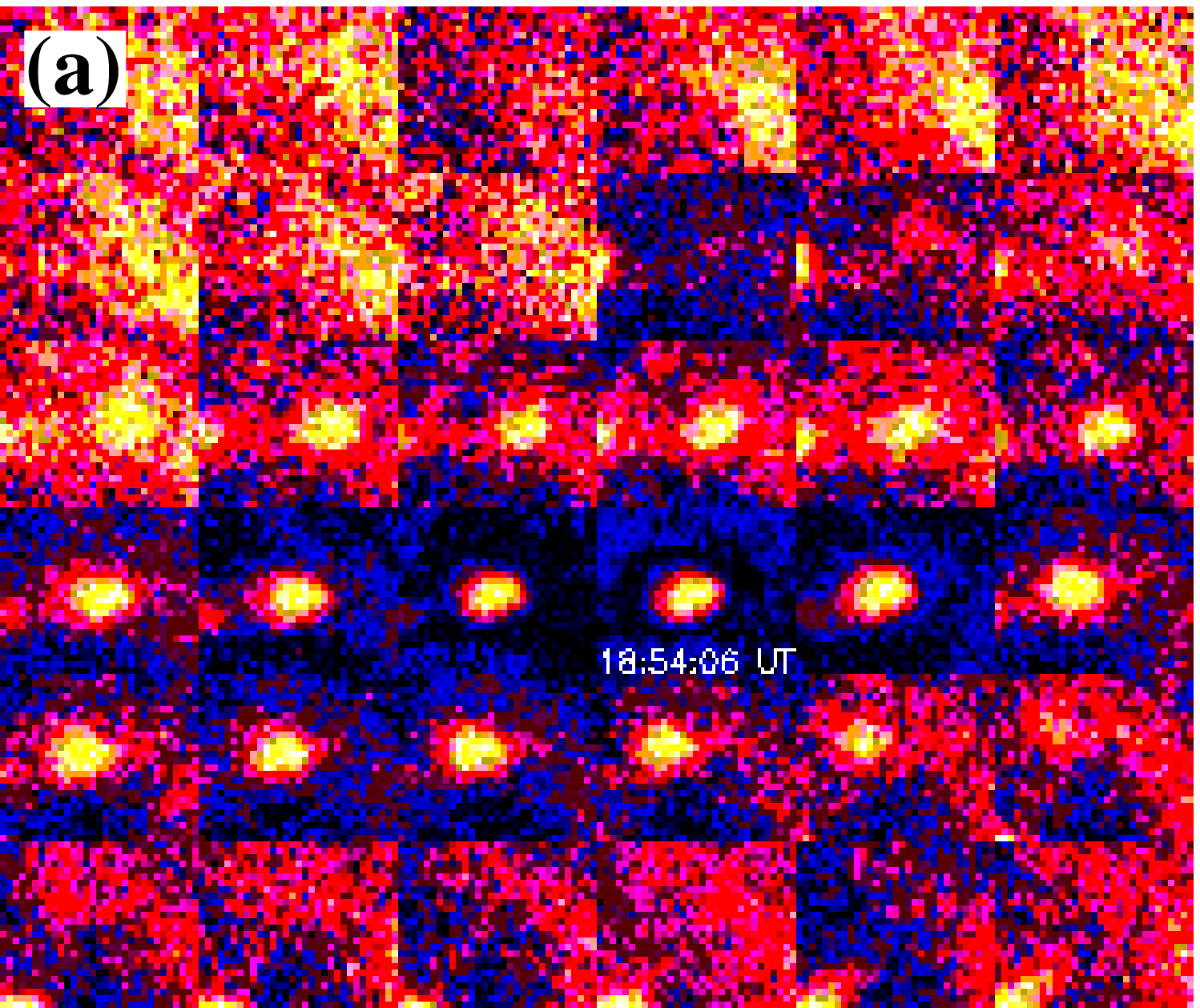}
\includegraphics[width=.215\linewidth, bb=0 0 360 360]{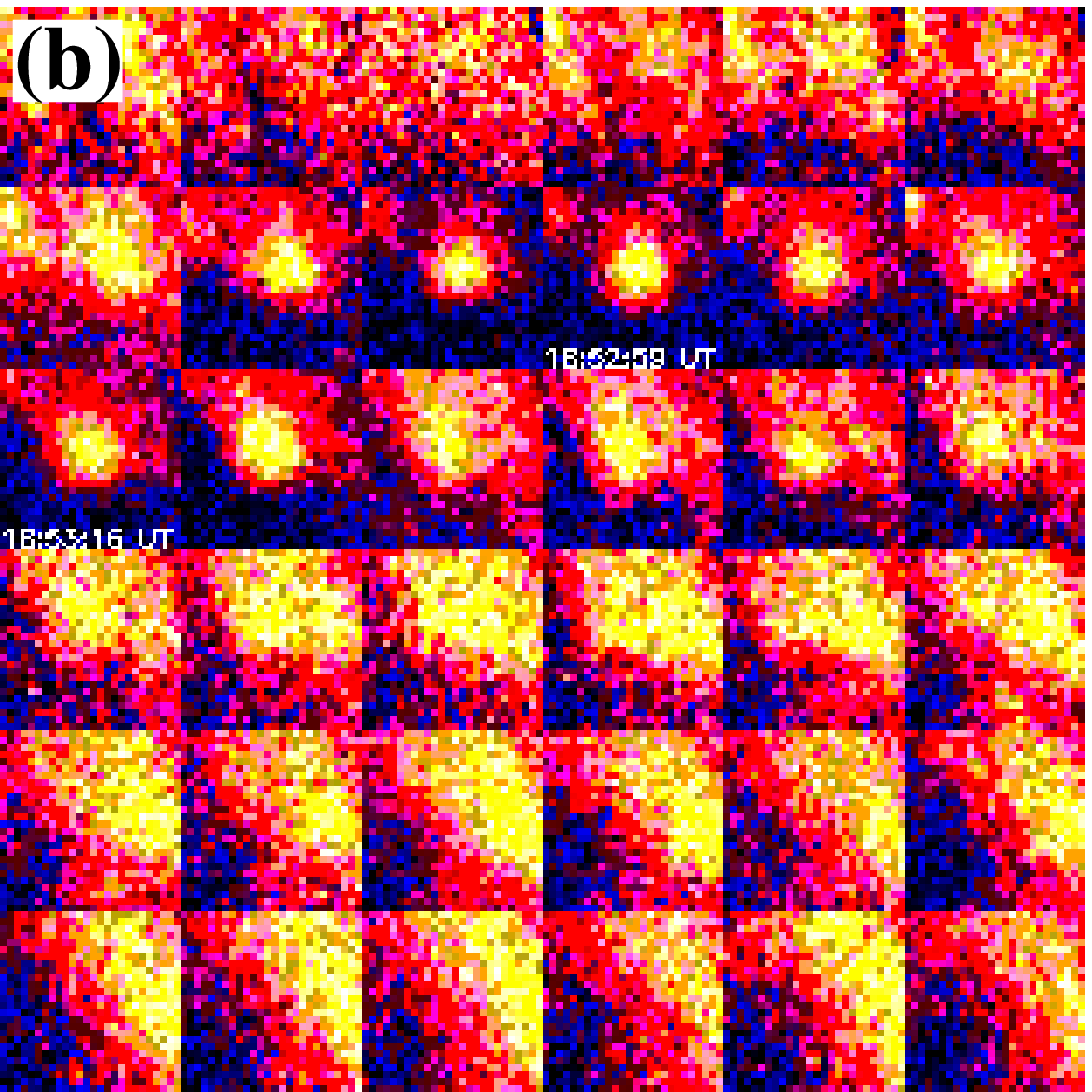}
\includegraphics[width=.215\linewidth, bb=0 0 360 360]{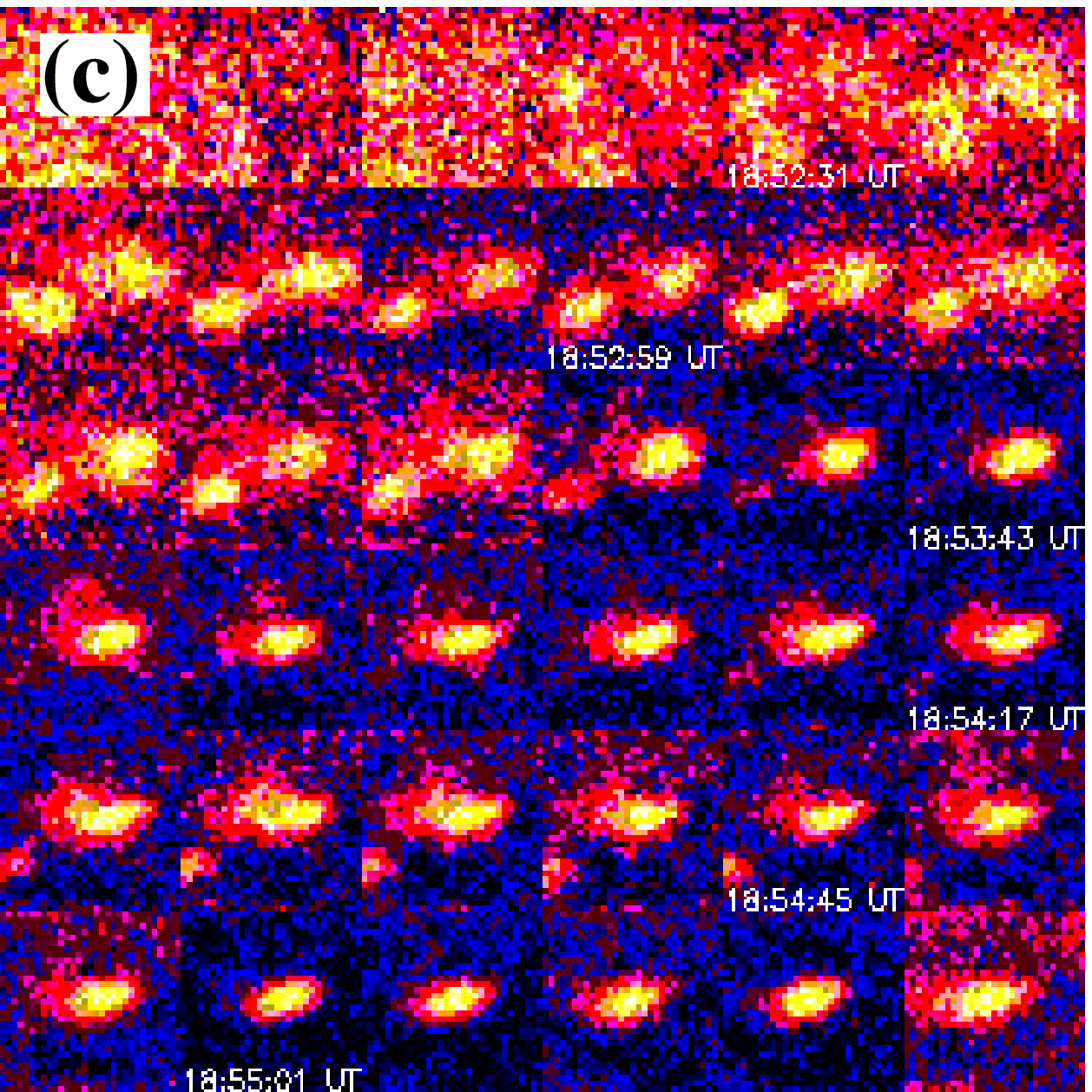}
\includegraphics[width=.266\linewidth, bb=0 0 446 360]{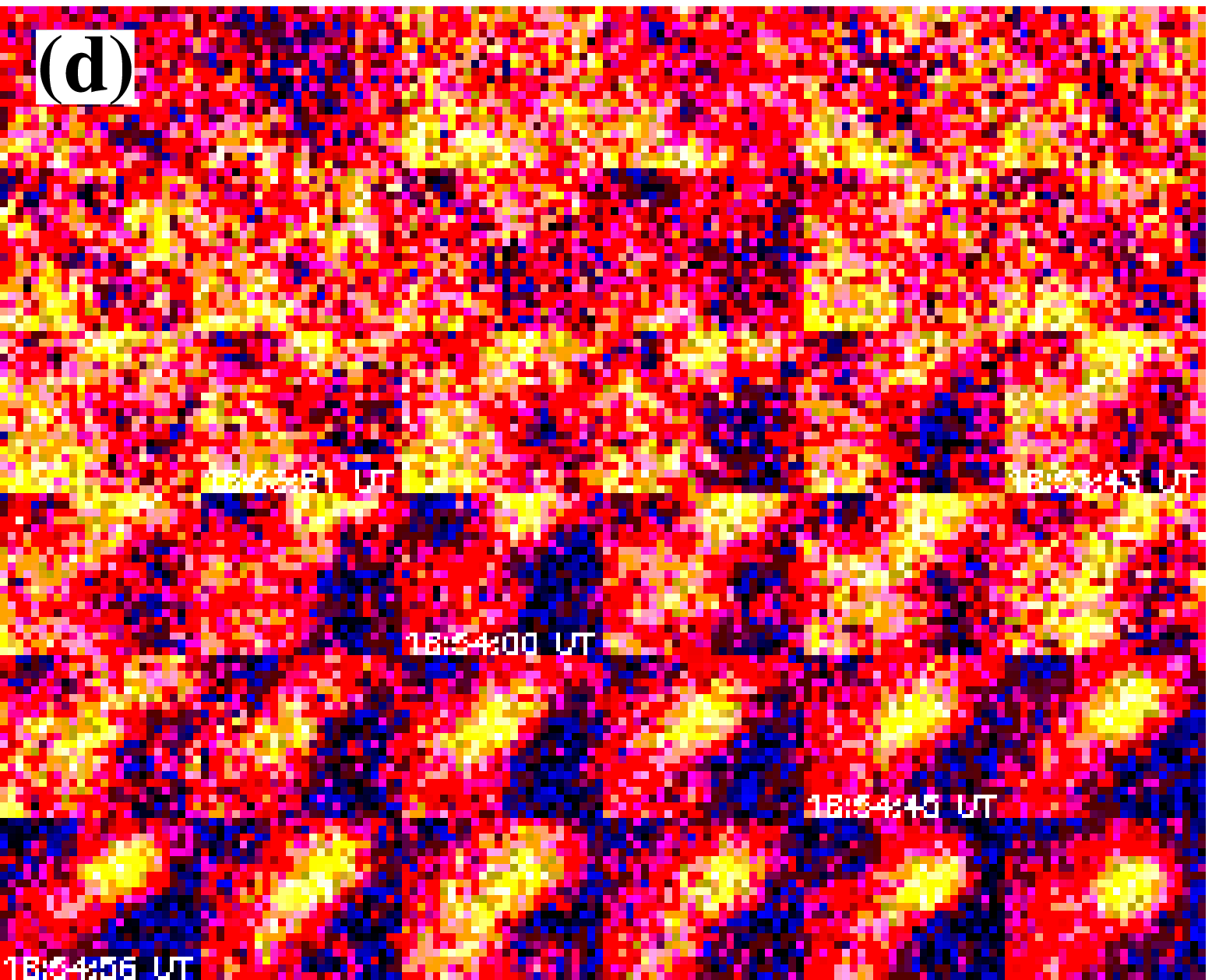}
\caption{Time evolution of the intensity associated with the four EBDs studied:
(a) EBD3 (FOV: 3.1\arcsec$\times$2.6\arcsec), (b) EBD4 (FOV:
2.6\arcsec$\times$2.6\arcsec), (c) EBD7 (FOV: 3.1\arcsec$\times$3.1\arcsec), 
and (d) EBD8 (FOV: 2.6\arcsec$\times$2.1\arcsec). Time increases from left to right, and top to
bottom. The time of the peaks marked on the light curves is annotated on the
images. The images have been normalised to the maximum EBD intensity in the Hi-C
time series.}
\label{fig:ts_ebd3}
\end{center}
\end{figure*}

In Fig.~\ref{fig:lc_ebd8}, we see a clear increase in intensity from 18:53:16 UT
to the end of the Hi-C time series: several peaks are observed during this
period. The identified peaks appear at 18:53:21 UT, 18:53:43 UT, 18:54:00 UT,
18:54:45 UT, and 18:54:56 UT (see also Fig.~\ref{fig:ts_ebd3}d). The peaks are
not all seen in the 193\AA\ SDO/AIA channel due to their short duration and the
time span between consecutive SDO/AIA images. The EBDs recur at a period of
about 20s. This short period and the increase of intensity suggest that the
energy released during the EBDS heats up the plasma without the possibility for
the conductive cooling to take place.   

\section{Magnetic Field Structure}
\label{sec:mag}

	\subsection{SDO/HMI Data and Extrapolated Magnetic Field Over Hi-C
	FOV} \label{sec:extrapol}

A line-of-sight magnetogram provided by SDO/HMI \citep{sch12} allows
us to extrapolate the potential magnetic field into the solar atmosphere. The
SDO/HMI magnetogram has been captured on July 11, 2012 at 18:54:00 UT with a
pixel size of 0.5\arcsec. 

\begin{figure}[!ht]
\begin{center}
\includegraphics[width=0.521\linewidth, bb= 0 0 410 360]{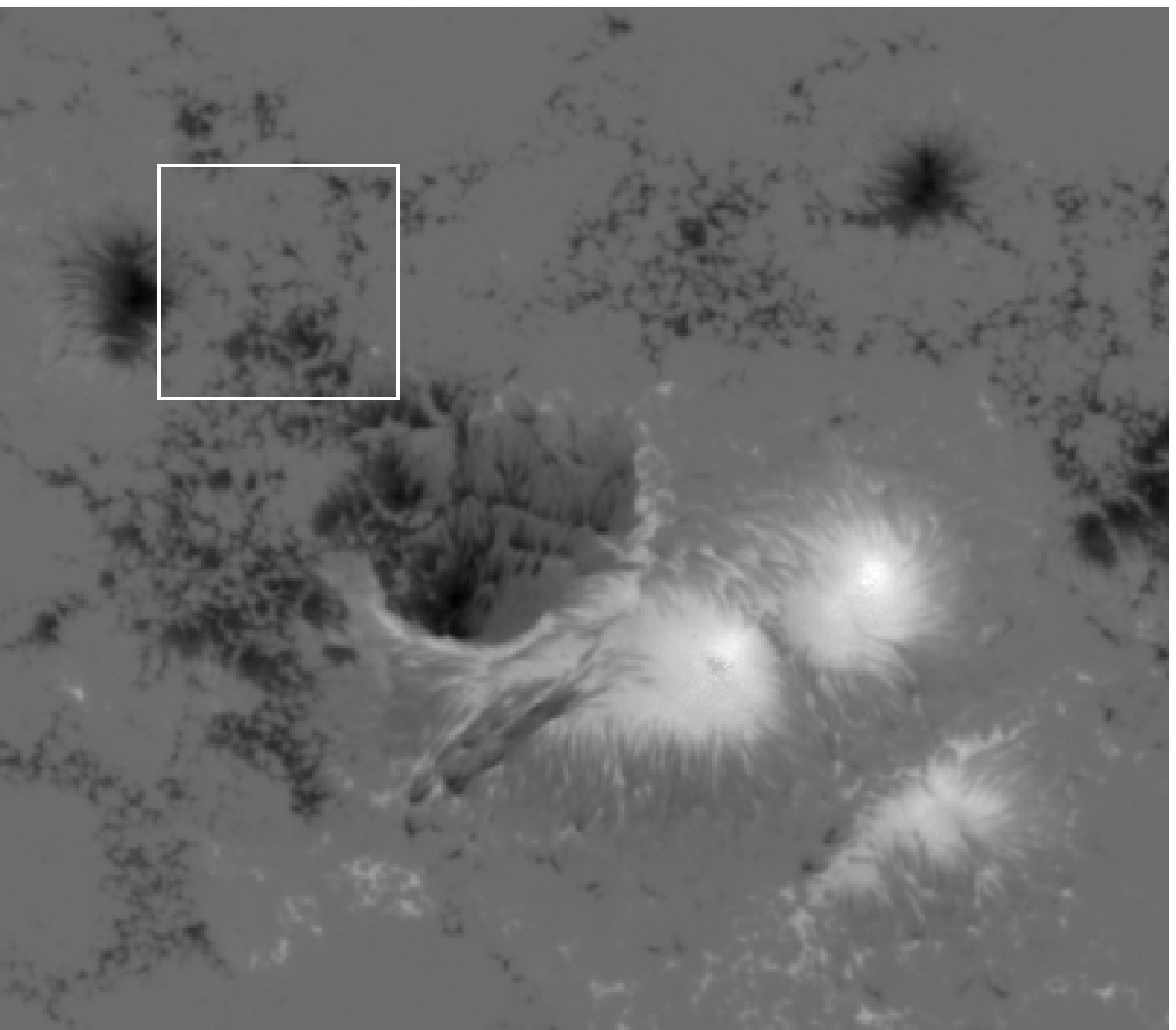}
\includegraphics[width=0.467\linewidth]{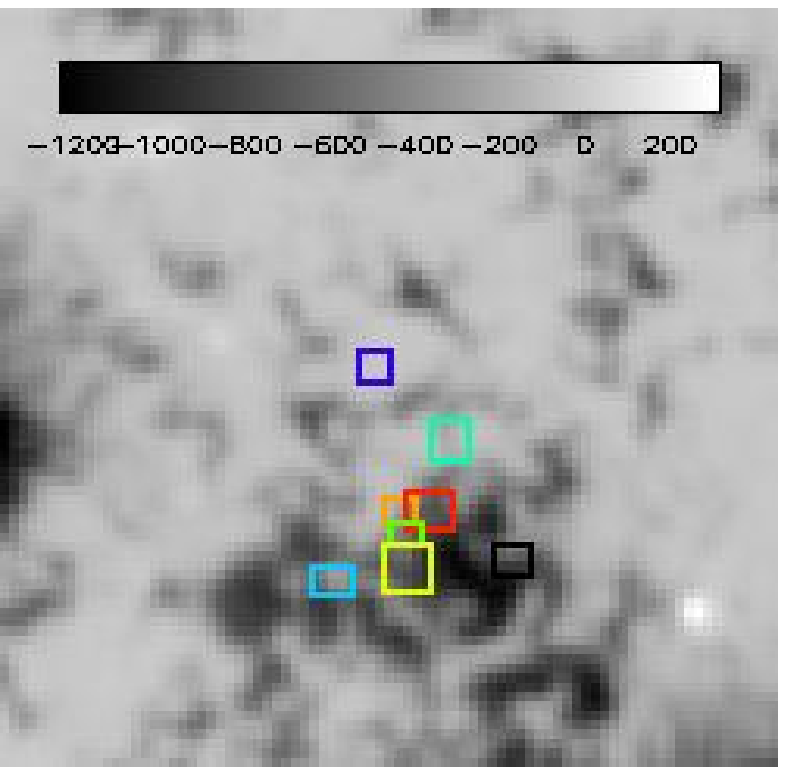}
\caption{(Left) Line-of-sight magnetic field distribution in the full HiC FOV
observed by SDO/HMI. The EBD FOV is indicated by the white rectangle; (Right)
close-up of the magnetic field in the area where the EBDs are observed (same
color coding as in Fig.~\ref{fig:hic_aia_int} bottom). The line-of-sight
magnetic field strength (G) is indicated by the color bar.}
\label{fig:bz}
\end{center}
\end{figure}

The potential field assumption considers that
\begin{equation}
\vec \nabla \times \vec B = \vec 0.
\end{equation}
The system of differential equations is solved as a boundary value problem in
defining the normal component of the magnetic field on each surface of the
computational box. Two types of potential fields are computed; (i) the potential
field for which the original SDO/HMI magnetogram (see Fig.~\ref{fig:bz} left) is
used as bottom boundary condition, and open boundary conditions on the sides and
top of the computational box are applied, (ii) the fully open potential field
within the computational box \cite[as defined by][]{aly84} for which the
magnetic field at the bottom boundary is unipolar. The first extrapolated
magnetic field corresponds to a minimum of magnetic energy for this set of
normal magnetic field distributions, whilst the second model corresponds to an
upper bound of magnetic energy for a force-free field \citep{aly84}. Thus the
difference between the magnetic energies derived from both models gives an upper
bound for the free magnetic energy contained in the magnetic configuration.

In Fig.~\ref{fig:bz} left, the selected SDO/HMI FOV can be seen
(144$\times$126 Mm$^{2}$), which includes most of AR11520 and for which the
total unsigned magnetic flux is nearly balanced (see Table~\ref{tab:bstrength}).
In Fig.~\ref{fig:bz} right, we plot the Hi-C FOV corresponding to the
location of the EBDs. Subsequently, in Table~\ref{tab:bstrength}, we summarize
the magnetic properties of the full FOV used to perform the potential field
extrapolation, and then reduced to the EBD FOV: 
\begin{itemize}
\item[-]{the minimum and maximum line-of-sight
magnetic field strength for the photospheric magnetogram ($B_{los}^{min}$,
$B_{los}^{max}$ respectively);} 
\item[-]{the total unsigned magnetic flux ($\phi_T$),
\begin{equation}
\phi_{T} = \int_S~|B_z|~dS
\end{equation}
where $S$ is the photospheric surface;}
\item[-]{the net magnetic flux ($\phi_{net}$),
\begin{equation}
\phi_{net} = \int_S~B_z~dS;
\end{equation}} 
\item[-]{the positive and negative magnetic fluxes
($\phi_+$, $\phi_-$ respectively):
\begin{equation}
\phi_{+} = \int_{\Omega^+}~B_z~dS \quad \quad \phi_{-} = \int_{\Omega^-}~B_z~dS
\end{equation}
where $\Omega^{+}$ and $\Omega^{-}$ are the domains of positive (resp. negative)
$B_z$ on the photospheric surface;}
\item[-]{the magnetic energy derived from
the potential magnetic field with open boundary conditions ($E_{pot}$) and from
the fully open magnetic field ($E_{open}$). By definition, the magnetic energy
$E_m$ for a magnetic field vector $B$ is given by
\begin{equation}
E_m = \int_{V}~\frac{B^2}{2\mu_0}~dV
\end{equation}
where $V$ is the volume of the computational box and $\mu_0$ is the vacuum
permeability.} 
\end{itemize}

\begin{figure}[!ht]
\begin{center}
\includegraphics[width=1.\linewidth]{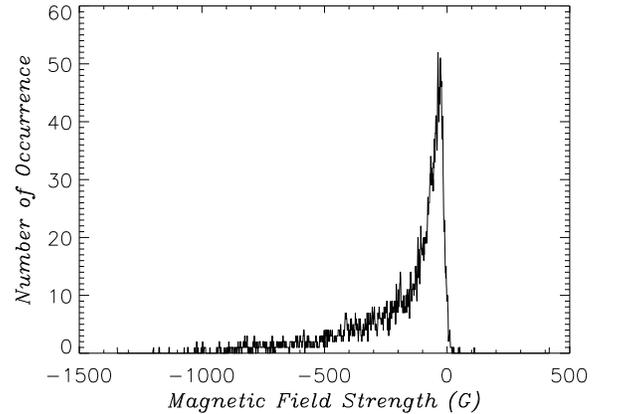}
\caption{Histogram of the line-of-sight magnetic field strength in the
region depicted in Fig.~\ref{fig:bz} right.}
\label{fig:histo}
\end{center}
\end{figure}

Note that the full FOV has a total unsigned magnetic flux through the
photosphere almost balanced, while the magnetic flux for the EBD FOV is
dominated by the negative flux. As can be seen in Fig.~\ref{fig:histo}, the
histogram of the line-of-sight magnetic field strength in the EBD area shows
that the negative polarity is dominant in this area and is (almost) unipolar.
However, it is worth noting that the negative polarity where the EBDs are
observed is not located in a sunspot (concentrated intense unipolar magnetic
field) but more like in a plage magnetic field region (diffuse magnetic field
which can contain parasitic polarities), so we cannot rule out the presence of
small-scale positive polarities, which cannot be resolved by SDO/HMI and which
can play an important role in terms of low-lying magnetic reconnection events.

\begin{table*}[!ht]
\begin{center}
\caption{Magnetic Properties}
\label{tab:bstrength}
\begin{tabular}{ccccccccc}
\tableline\tableline \\[-0.2cm]
 & $B^{min}_{los}$ & $B^{max}_{los}$ & $\phi_T$ & $\phi_{net}$ & $\phi_+$ 
 & $\phi_-$ & $E_{pot}$ & $E_{open}$ \\
 & (G) & (G) & (Mx) & (Mx) & (Mx) & (Mx) & (erg) & (erg) \\[0.1cm]
\tableline \\[-0.2cm]
Full FOV & -1830 & 2463 & 5.97$\times$10$^{22}$& -9.00$\times$10$^{21}$&
2.53$\times$10$^{22}$& -3.43$\times$10$^{22}$ & 1.68$\times$10$^{33}$ &
3.03$\times$10$^{33}$ \\
EBD FOV & -1232 & 316 & 2.61$\times$10$^{21}$ & -2.60$\times$10$^{21}$ & 7.43$\times$10$^{18}$ &
-2.60$\times$10$^{21}$ & 5.20$\times$10$^{31}$ & 8.98$\times$10$^{31}$ \\
\tableline
\end{tabular}
\end{center}
\end{table*}

\begin{figure}[!ht]
\begin{center}
\includegraphics[width=0.9\linewidth]{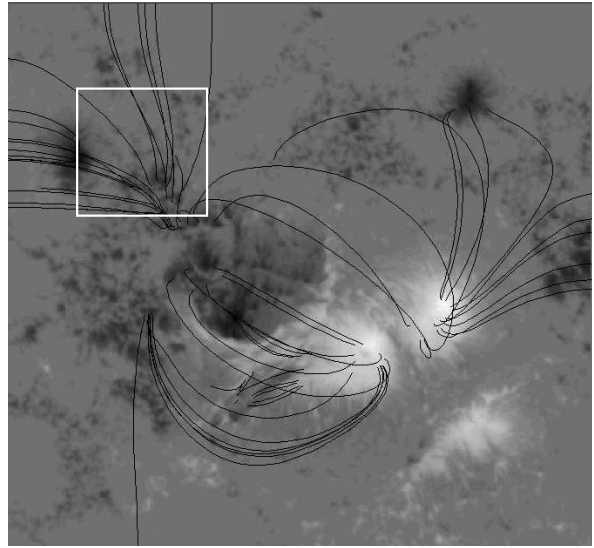}
\caption{Selection of magnetic field lines within AR11520 extracted from the
potential field extrapolation. The background image is the line-of-sight
magnetic field observed by SDO/HMI. The white box indicates the field-of-view
containing the EBDs as in Fig.~\ref{fig:bz}.}
\label{fig:pot_full}
\end{center}
\end{figure}

In Fig.~\ref{fig:pot_full}, we plot a selection of magnetic field lines, which
describes the global magnetic geometry of AR11520 derived from a potential field
approximation. Fig.~\ref{fig:pot_full} highlights the magnetic connectivity
between the sunspots of AR11520 and outside (West side) towards AR11519 and
AR11521. We also plot a significant amount of magnetic field lines in and near
the area containing the EBDs: this shows that those magnetic field lines are
open (i.e., leaving the computational box) towards the East and North sides. It
is worth noting that the magnetic field lines oriented towards the North are
associated with transequatorial loops connecting to the positive polarity of a
diffuse, decaying active region in the Northern hemisphere. 

Table~\ref{tab:bstrength} displays the magnetic energy of the potential and open
magnetic field configuration for both the full FOV and EBD FOV. The magnetic
energy of the full FOV is typical for an active region with a photospheric flux
of the order of 10$^{22}$ Mx. The upper bound for the free magnetic energy is
estimated to be $E_{open} - E_{pot} = 1.35\times10^{33}$ erg, consistent with
the high-level of activity in AR11520 for a couple of days around the Hi-C
observations (including an X1.4 flare on 12 July 2012). The upper bound for the
free magnetic energy in the EBD FOV is estimated to 3.78$\times$ 10$^{31}$ erg.
Therefore, energy-release events such as nanoflare (with an energy of $10^{24}$
erg) can be generated within the EBD FOV.  

	\subsection{EBD's Magnetic Field Structure} \label{sec:ebd_fl}

\begin{figure}[!ht]
\begin{center}
\includegraphics[width=.385\linewidth]{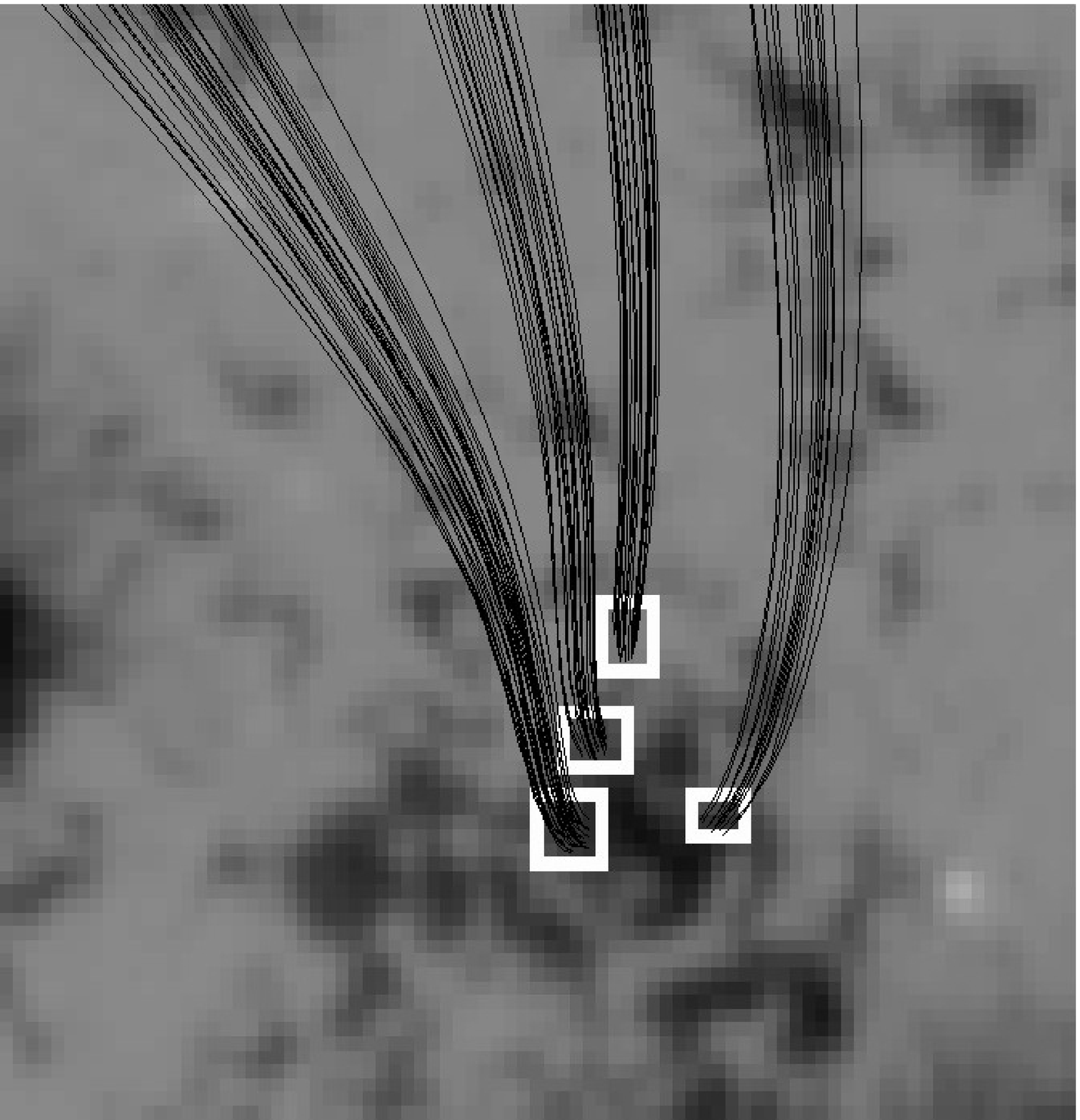}
\includegraphics[width=.585\linewidth]{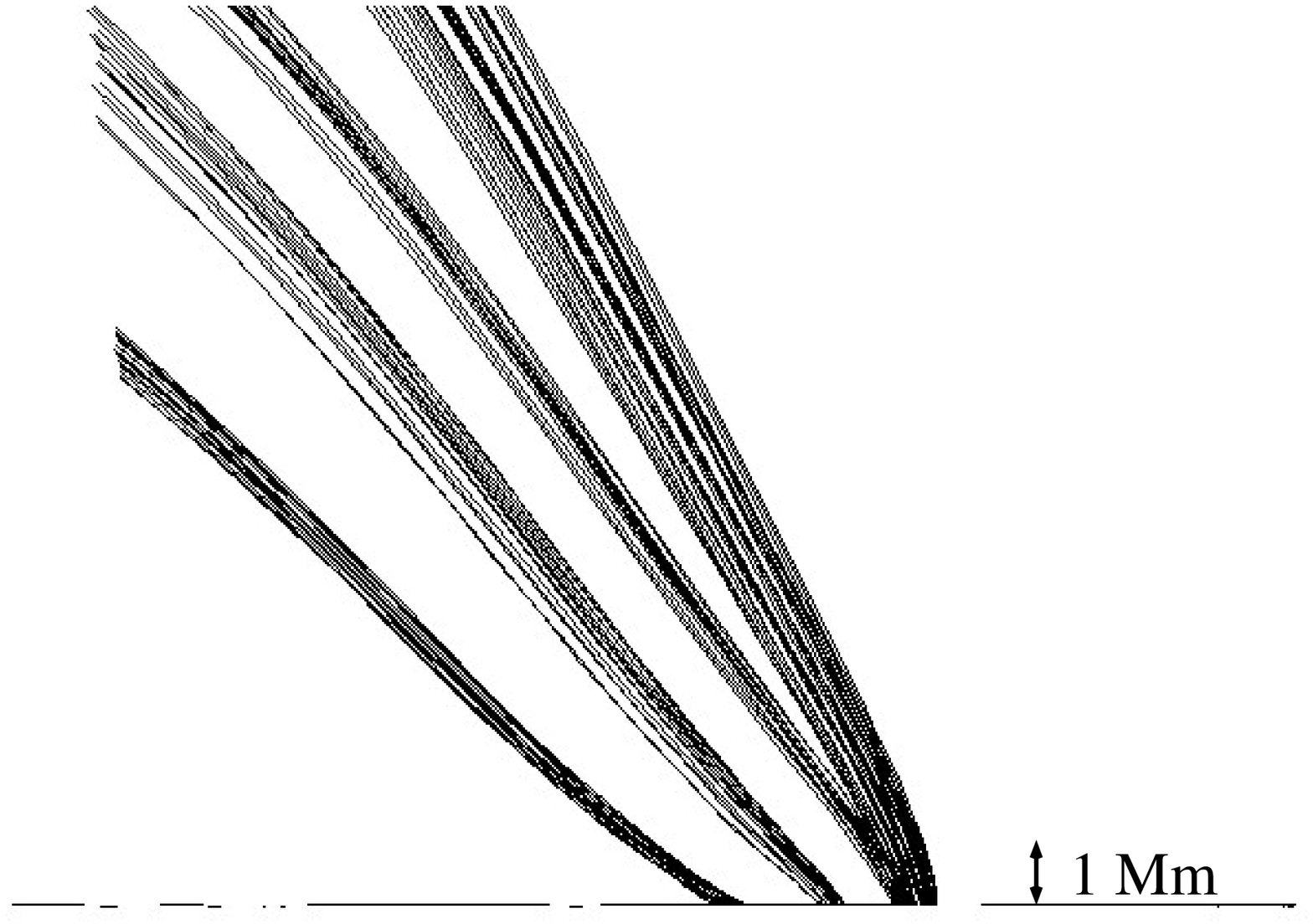}
\caption{A selection of magnetic field lines for the four EBDs studied in
Section~\ref{sec:ebd_fl}. (Left) top view with the EBD's location depicted by
white rectangles (see Fig.~\ref{fig:hic_aia_int}); (Right) view from the East
side (North is on the left-hand side)}
\label{fig:fl_ebd}
\end{center}
\end{figure}

We now study the magnetic properties of individual EBDs considering the same
four events as in the Section~\ref{sec:dyn}. For the sake of clarity, in
Fig.~\ref{fig:fl_ebd}, we only plot a selection of magnetic field lines associated
with  EBD3 (Cat I), EBD4 (Cat II), EBD7 (Cat III), and EBD8 (Cat IV). For these
EBDs, the magnetic field lines are leaving the computation box towards the
North-East consistent with the observed transequatorial loops connecting two
active regions (see Section~\ref{sec:extrapol}). We extract the magnetic information
for each set of magnetic field lines: it is above 1 Mm that the magnetic field
strength becomes a smoothly, decaying function of height. Below 1 Mm, the
magnetic field strength is influenced by the complexity of the magnetic field
and the change in orientation of the transverse component (parallel to the
photosphere). It is also below 1 Mm that the magnetic field lines plotted in
Fig.~\ref{fig:fl_ebd} can be considered as radial. The magnetic field strength
of the four EBDs below 1 Mm varies between 200 G and 280 G. 


We estimate the magnetic energy associated with the EBDs by considering the
potential field within a sphere of diameter 0.68 Mm (characteristic size of
EBDs) and within 1Mm above the photosphere where the magnetic field strength has
been estimated between 200 and 280 G. The magnetic energy is thus between
1.98$\times$10$^{26}$ and 3.87$\times$10$^{26}$ erg. An upper bound for the free
magnetic energy is about 80\% of the previous values, i.e.,
1.58--3.1$\times$10$^{26}$ erg.

The free magnetic energy estimates are of 10$^{26}$ erg which are similar to the
energy associated with a single micro-flare. Thus the magnetic field associated
with the EBDs contains enough free magnetic energy to power single or multiple
nanoflares as envisioned by \cite{par88}, knowing that a small part of the free
magnetic energy is commonly released during flares \cite[e.g.,][]{ama00}. By
invoking an energy release site reduced by a factor of 3 (0.23 Mm instead of
0.68 Mm), the magnetic energy estimated reaches the nanoflare scale. This
magnetic energy estimate evidences that the nanoflare model can be a viable
mechanism to release magnetic energy at low-heights in the solar atmosphere and
thus heat the coronal part of the loop. 

\section{Conclusions}
\label{sec:concl}

The high resolution images obtained in the EUV by Hi-C allow us to determine the
smallest brightenings observed to date in an active region in extreme
ultra-violet light, called EUV bright dots (EBDs). These EBDs sparkle at the
footpoints of large-scale coronal loops. We have grouped the EBDs into four
categories depending on their recurrence. Except for the fourth category
(successive, short-lived EBDs), individual EBDs which can be extracted from the
Hi-C time series have the same characteristic duration (about 25s) and length
scale (0.68 Mm). Using a potential field extrapolation of the SDO/HMI
line-of-sight photospheric magnetic field, we have determined that the EBDs
observed in the Hi-C FOV are located at the footpoints of large-scale magnetic
loops connecting the active regions 11520 to a decaying active region in the
Northern hemisphere. The above properties suggest that EBDs might be a signature
of an impulsive release of energy located at the base of coronal loops. 

In terms of energetics, the magnetic energy estimate is of the order of
10$^{26}$ erg which is 100 times more than required by the nanoflare model
suggested by \citet{par88}. However, our estimate is an upper bound for the free
magnetic energy than can be released, thus suggesting that smaller scale
structures can exist. Nevertheless, \citet{tes13} and \citet{win13} have
estimated the radiated energy to 10$^{24}$ erg, which is compatible with the
order of magnitude of the magnetic energy: according to \citet{ems12} who
performed a complete study of the energy partition in eruptive events, the
magnetic energy budget can be two orders of magnitude larger than the radiated
energy.

Even if it is not possible to compare quantitatively due to the lack of
spectroscopic data overlying the Hi-C FOV, it is important to mention
the observations of \citet{har09}, which showed that small-scale brightenings at
the footpoints of coronal loops can be observed with a strong redshift velocity
close to the local sound speed. \citet{har09} suggested that spatially
unresolved ($<$720 km) structures could exist at the footpoint of coronal loops.

According to \cite{walsh95}, the conduction time-scale can be expressed as
$\tau_c = pL^2 / \kappa_0T^{7/2}$ where $p$ is the plasma pressure, $L$ the
characteristic length, $\kappa_0$ the thermal conductivity in the corona
($\kappa_0 = 10^{-11}$), and $T$ is the temperature. The characteristic length
$L$ of the loop is chosen to be the loop length along which the heating is
assumed to take place: $L$ = 5 Mm. For a plasma pressure of 0.1
dyne$\cdot$cm$^{-2}$ and a temperature range from 0.5 to 1.5 MK, the conductive
time-scale is between 6 s at 1.5 MK and 5 min at 0.5 MK. A characteristic time
of 25 s gives a characteristic temperature of 1 MK. Thus, the observed life-time
of EBDs maybe compatible with the properties of a coronal plasma. 

With the restricted observations of the Hi-C instrument (short flight and
limited FOV), questions arise for
further study of these elementary events: how often/common are EBDs in active
regions? Are they always located at the footpoints of long coronal loops? What
is their contribution towards the global heating of the corona? Can EBDs be
associated with Type 2 spicules or even Ellerman bombs, both having similar time
scales? Several of these
questions can be addressed by finding criteria which will enable the detection
of EBDs in SDO/AIA images going forward, and by using high resolution
chromospheric spectroscopic observations from IRIS.

\acknowledgments
We thank Jonathan Cirtain and Amy Winebarger for their help and discussion on
the Hi-C instrument and to improve this manuscript. MSFC/NASA led the mission
and partners include the Smithsonian Astrophysical Observatory in Cambridge,
Mass.; Lockheed Martin's Solar Astrophysical Laboratory in Palo Alto, Calif.;
the University of Central Lancashire in Lancashire, England; and the Lebedev
Physical Institute of the Russian Academy of Sciences in Moscow.



\begin{thebibliography}{}

\bibitem[Alexander et al.(2013)]{ale13}
	{Alexander}, C. E., {Walsh}, R. W., {R{\'e}gnier}, S., et al. 2013,
	\apjl, 775, L32 
	
\bibitem[Aly(1984)]{aly84}
	{Aly}, J.~J. 1984, \apj, 283, 349

\bibitem[Amari et al.(2000)]{ama00}
	{Amari}, T., {Luciani}, J.~F., {Mikic}, Z., {Linker}, J. 2000,
	\apjl, 529, L49

\bibitem[Beck et al.(2007)]{bec07}
	{Beck}, C., {Bellot Rubio}, L.~R., {Schlichenmaier}, R., 
	{S{\"u}tterlin}, P. 2007, \aap, 472, 607
	
\bibitem[Berger et al.(2004)]{ber04}
	{Berger}, T.~E., {Rouppe van der Voort}, L.~H.~M., {L{\"o}fdahl},
	M.~G., {Carlsson}, M., {Fossum}, A., {Hansteen}, V.~H.,
	{Marthinussen}, E.,  {Title}, A., {Scharmer}, G. 2004, \aap, 428, 613

\bibitem[Boerner et al.(2012)]{boe12}
	{Boerner}, P., {Edwards}, C., {Lemen}, J., {Rausch}, A., 
	{Schrijver}, C., {Shine}, R., {Shing}, L., {Stern}, R.,
	{Tarbell}, T., {Title}, A., {Wolfson}, C.~J., {Soufli}, R., 
	{Spiller}, E., {Gullikson}, E., {McKenzie}, D., {Windt}, D., 
	{Golub}, L., {Podgorski}, W., {Testa}, P., {Weber}, M. 2012, 
	\solphys, 275, 41

\bibitem[Cirtain et al.(2013)]{cirt13}
	{Cirtain}, J.~W., {Golub}, L., {Winebarger}, A.~R., {DePontieu}, B.,
	{Kobayashi}, K., {Moore}, R.~L., {Walsh}, R.~W., {Korreck}, K.~E.,
	{McCauley}, P., {Title}, A., {Kuzin}, S., {DeForest}, C.~E. 2013,
	Nature, 493, 501 
	
\bibitem[del Zanna et al.(2002)]{del02}
	{del Zanna}, G., {Landini}, M., {Mason}, H.~E. 2002, 385, 968

\bibitem[Emslie et al.(2012)]{ems12}
	{Emslie}, A.~G., {Dennis}, B.~R., {Shih}, A.~Y., {Chamberlin}, P.~C., 
	{Mewaldt}, R.~A., {Moore}, C.~S., {Share}, G.~H., {Vourlidas}, A., 
	{Welsch}, B.~T. 2012, \apj, 759, 71

\bibitem[Guennou et al.(2012)]{gue12}
	{Guennou}, C., {Auch{\`e}re}, F., {Soubri{\'e}}, E., 
	{Bocchialini}, K.,{Parenti}, S., {Barbey}, N. 2012, \apjs, 203, 26

\bibitem[Hara(2009)]{har09} {Hara}, H. 2009, in The Second Hinode Science
	Meeting: Beyond Discovery-Toward Understanding, ASPC, eds. {Lites}, B.,
	{Cheung}, M., {Magara}, T., {Mariska}, J., {Reeves}, K., 415, 252

\bibitem[Jordan et al.(1987)]{jor87}
	{Jordan}, C., {Ayres}, T.~R., {Brown}, A., {Linsky}, J.~L., 
	{Simon}, T. 1987, \mnras, 225, 903

\bibitem[Klimchuk(2006)]{klim06}
	{Klimchuk}, J.~A. 2006, \solphys, 234, 41

\bibitem[Kobayashi et al.(2013)]{kob13}
	Kobayashi, K., Cirtain, J., Winebarger, A.~R., Korreck, K.~E., Golub,
	L., Walsh, R.~W., De Pontieu, B., DeForest, C.~E., Title, A., Kuzin, S.,
	Savage, S., Beabout, D., Beabout, B., Podgorski, W., Caldwell, D.,
	McCracken, K., Ordway, M., Bergner, H., Gates, R., McKillop, S.,
	Cheimets, P., Platt, S., Mitchell, N., Windt, D. 2013, submitted to
	\solphys
	 
\bibitem[Lemen et al.(2012)]{lemen12}
	{Lemen}, J.~R., {Title}, A.~M., {Akin}, D.~J., {Boerner}, P.~F., 
	{Chou}, C., et al.
	2012, \solphys, 275, 17

\bibitem[Mackay et al.(2000)]{mac00}
	{Mackay}, D.~H., {Galsgaard}, K., {Priest}, E.~R., 
	{Foley}, C.~R. 2000, \solphys, 193, 93


\bibitem[O'Dwyer et al.(2010)]{odw10}
	{O'Dwyer}, B., {Del Zanna}, G., {Mason}, H.~E., {Weber}, M.~A., 
	{Tripathi}, D. 2010, \aap, 521, 21

\bibitem[Parker(1988)]{par88}
	{Parker}, E.~N. 1988, \apj, 330, 474

\bibitem[Pesnell, Thompson and Chamberlin(2012)]{pes12}
	{Pesnell}, W.~D. and {Thompson}, B.~J. and {Chamberlin}, P.~C. 2012,
	\solphys, 275, 3
	
\bibitem[Priest et al.(2000)]{pri00}
	{Priest}, E.~R., {Foley}, C.~R., {Heyvaerts}, J., {Arber}, T.~D., 
	{Mackay}, D., {Culhane}, J.~L., {Acton}, L.~W. 2000, \apj, 539, 1002

\bibitem[Reale(2002)]{reale02}
	{Reale}, F. 2002, \apj, 580, 566

\bibitem[Reale(2010)]{reale10}
	{Reale}, F. 2010, Living Reviews in Solar Physics, 7, 5

\bibitem[R{\'e}gnier and Priest(2007)]{reg07}
	{R{\'e}gnier}, S. and {Priest}, E.~R. 2007, \aap, 468, 701

\bibitem[Rouppe van der Voort et al.(2005)]{rou05}
	{Rouppe van der Voort}, L.~H.~M., {Hansteen}, V.~H., {Carlsson}, M., 
	{Fossum}, A., {Marthinussen}, E., {van Noort}, M.~J., 
	{Berger}, T.~E. 2005, \aap, 435, 327

\bibitem[Rutten et al.(2004)]{rut04}
	{Rutten}, R.~J., {Hammerschlag}, R.~H., {Bettonvil}, F.~C.~M., 
	{S{\"u}tterlin}, P., {de Wijn}, A.~G. 2004, \aap, 413, 1183
	
\bibitem[Scharmer(2003)]{sch03} {Scharmer}, G.~B. and {Bjelksjo}, K. and
	{Korhonen}, T.~K. and  {Lindberg}, B. and {Petterson}, B. 2003, Society
	of Photo-Optical Instrumentation Engineers (SPIE) Conference Series,
	 eds {{Keil}, S.~L. and {Avakyan}, S.~V.}, 4853, 341
	 
\bibitem[Scherrer et al.(2012)]{sch12}
	{Scherrer}, P.~H., {Schou}, J., {Bush}, R.~I., {Kosovichev}, A.~G., 
	{Bogart}, R.~S., {Hoeksema}, J.~T., {Liu}, Y., {Duvall}, T.~L., 
	{Zhao}, J., {Title}, A.~M., {Schrijver}, C.~J., {Tarbell}, T.~D., 
	{Tomczyk}, S. 2012, \solphys, 275, 207

\bibitem[Testa et al.(2013)]{tes13}
	{Testa}, P., {De Pontieu}, B., {Mart{\'{\i}}nez-Sykora}, J., 
	{DeLuca}, E., {Hansteen}, V., {Cirtain}, J., {Winebarger}, A., 
	{Golub}, L., {Kobayashi}, K., {Korreck}, K., {Kuzin}, S., 
	{Walsh}, R., {DeForest}, C., {Title}, A., {Weber}, M. 2013, \apjl, 770, 
	L1

\bibitem[Viall and Klimchuk(2011)]{via11}
	{Viall}, N.~M. and {Klimchuk}, J.~A. 2011, \apj, 738, 24

\bibitem[Viall and Klimchuk(2012)]{via12}
	{Viall}, N.~M. and {Klimchuk}, J.~A. 2012, \apj, 753, 35

\bibitem[Walsh, Bell and Hood(1995)]{walsh95}
	{Walsh}, R.~W., {Bell}, G.~E., {Hood}, A.~W. 1995, \solphys, 161, 83

\bibitem[Winebarger et al.(2013)]{win13}
	{Winebarger}, A.~R., {Walsh}, R.~W., {Moore}, R., {De Pontieu}, B.,
	{Hansteen}, V., {Cirtain}, J., {Golub}, L., {Kobayashi}, K., {Korreck}, K.,
	{DeForest}, C., {Weber}, M., {Title}, A., {Kuzin}, S. 2013, \apj, 771,
	21

\bibitem[Winebarger et al.(2003)]{win03}
	{Winebarger}, A.~R., {Warren}, H.~P., {Seaton}, D.~B. 2003, 
	\apj, 593, 1164

\end{thebibliography}
\end{document}